\documentclass{IEEEtran4PSCC}
\ifCLASSINFOpdf
   \usepackage[pdftex]{graphicx}
\else
   \usepackage[dvips]{graphicx}
\fi
\usepackage{amsmath,amsfonts}
\usepackage{amsthm}
\usepackage{amssymb}
\usepackage{csquotes}
\usepackage{blkarray}
\usepackage{bm}
\usepackage{cases}
\usepackage{xcolor}
\usepackage{hyperref}
\usepackage{booktabs}
\usepackage{multirow}
\usepackage{setspace}
\usepackage{enumitem} 
\usepackage{fancybox}
\usepackage{tcolorbox}
\usepackage{algorithm}
\usepackage[noend]{algpseudocode}

\usepackage{array}
\usepackage[caption=false,font=normalsize,labelfont=sf,textfont=sf]{subfig}
\usepackage{textcomp}
\usepackage{stfloats}
\usepackage{url}
\usepackage{verbatim}
\usepackage{comment}
\usepackage{graphicx}
\allowdisplaybreaks
\newtheorem{theorem}{Theorem}

\usepackage{caption}
\usepackage[noadjust]{cite}
\usepackage[clock]{ifsym}

\usepackage{float}
\floatstyle{ruled}
\newfloat{model}{thp}{lop}
\floatname{model}{Model}

\newfloat{projection}{thp}{lop}
\floatname{projection}{Projection}

\definecolor{niceblue}{rgb}{0, 0.5, 1.0}
\definecolor{niceblue}{rgb}{0.125, 0.406, 0.852}

\hypersetup{
    colorlinks=true,
    linkcolor=niceblue,
    filecolor=magenta,      
    urlcolor=niceblue,
    citecolor=niceblue}
\usepackage{graphicx}
\usepackage{textcomp}
\newcommand{\newtxt}[1]{{#1}}
\newcommand{\btt}[1]{{\fontfamily{lmtt}\selectfont #1}}

\usepackage{balance}

\makeatletter
\let\old@ps@headings\ps@headings
\let\old@ps@IEEEtitlepagestyle\ps@IEEEtitlepagestyle
\def\psccfooter#1{%
    \def\ps@headings{%
        \old@ps@headings%
        \def\@oddfoot{\strut\hfill#1\hfill\strut}%
        \def\@evenfoot{\strut\hfill#1\hfill\strut}%
    }%
    \def\ps@IEEEtitlepagestyle{%
        \old@ps@IEEEtitlepagestyle%
        \def\@oddfoot{\strut\hfill#1\hfill\strut}%
        \def\@evenfoot{\strut\hfill#1\hfill\strut}%
    }%
    \ps@headings%
}
\makeatother

\psccfooter{%
        \parbox{\textwidth}{\hrulefill \\ \small{23rd Power Systems Computation Conference} \hfill \begin{minipage}{0.2\textwidth}\centering \vspace*{4pt} \includegraphics[scale=0.06]{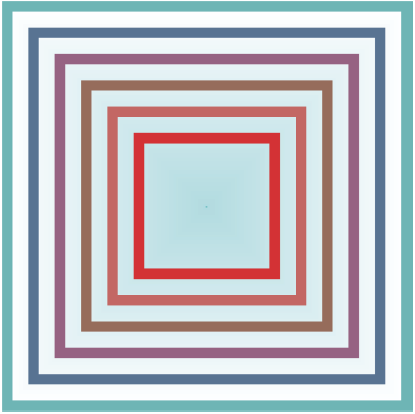}\\\small{PSCC 2024} \end{minipage} \hfill \small{Paris, France --- June 4 -- 7, 2024}}%
}

\begin{document}
%


\title{A Parallelized, Adam-Based Solver for Reserve and Security Constrained AC Unit Commitment}

\author{
\IEEEauthorblockN{Samuel Chevalier}
\IEEEauthorblockA{Department of Electrical Engineering \\
University of Vermont\\
Burlington, Vermont, USA\\
schevali@uvm.edu}
}


\maketitle

\begin{abstract}
Power system optimization problems which include the nonlinear AC power flow equations require powerful and robust numerical solution algorithms. Within this sub-field of nonlinear optimization, interior point methods have come to dominate the solver landscape. Over the last decade, however, a number of efficient numerical optimizers have emerged from the field of Machine Learning (ML). One algorithm in particular, Adam, has become the optimizer-of-choice for a massive percentage of ML training problems (including, e.g., the training of GPT-3), solving some of the largest unconstrained optimization problems ever conceived of. Inspired by such progress, this paper designs a parallelized Adam-based numerical solver to overcome one of the most challenging power system optimization problems: security and reserve constrained AC Unit Commitment. The resulting solver, termed \btt{QuasiGrad}, recently competed in the third ARPA-E Grid Optimization (GO3) competition. In the day-ahead market clearing category (with systems ranging from 3 to 23,643 buses over 48 time periods), \btt{QuasiGrad}'s aggregated market surplus scores were within 5$\%$ of the winningest market surplus scores. The \btt{QuasiGrad} solver is now released as an open-source Julia package: \btt{QuasiGrad.jl}. The internal gradient-based solver (Adam) can easily be substituted for other ML-inspired solvers (e.g., AdaGrad, AdaDelta, RMSProp, etc.). Test results from large experiments are provided.

\end{abstract}

\begin{IEEEkeywords}
AC unit commitment, Adam, optimal power flow, market surplus, mixed-integer, security constraints
\end{IEEEkeywords}

\thanksto{\noindent Submitted to the 23rd Power Systems Computation Conference (PSCC 2024).\\
This work was supported by the HORIZON-MSCA-2021 Postdoc Fellowship Program, Project \#101066991 -- TRUST-ML.}

\section{Introduction}
\IEEEPARstart{T}{he} first two \newtxt{Advanced Research Projects Agency–Energy} (ARPA-E) Grid Optimization competitions (GO1, GO2) focused on various flavors of the Security Constrained Optimal Power Flow (SCOPF) problems~\cite{aravena2023recent,GO2_review,GO2}. Within these competitions, the most successful numerical solution techniques leveraged interior point solvers~\cite{aravena2023recent}. The winningest approach in GO2, for example, used Ipopt via the Gravity modeling framework~\cite{hijazi2018gravity}. The third Grid Optimization competition (GO3), which recently concluded, focused on multi-period dynamic markets. More specifically, the GO3 market clearing problem incorporated security and reserve constrained AC unit commitment (ACUC) with topology optimization, and it asked competitors to maximize a social surplus function within the context of real-time, day-ahead, and week-look-ahead markets.

\begin{figure}
\centering
\includegraphics[width=1\columnwidth]{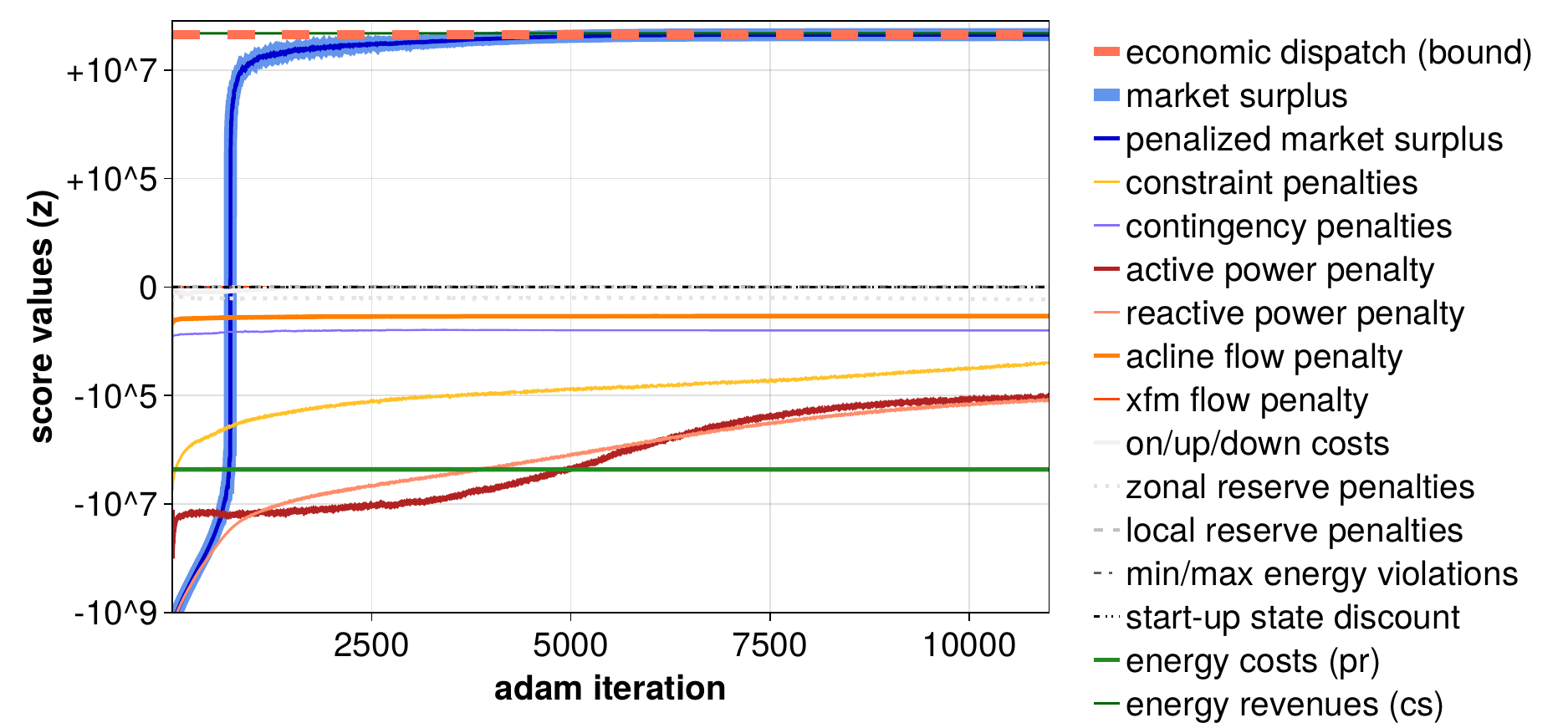}
\caption{Illustrated is an Adam solve on a 617-bus, 18 time period, real-time market clearing test case (integers relaxed); this system is initialized with a copper pate economic dispatch solution (LP), whose upper bound is given as the orange dashed line. Within several thousand iterations, Adam finds an AC network solution to within 1\% of this global bound. A single back-propagation (i.e., gradient calculation) through this entire system, include all $18\times562$ contingencies, takes $\sim$24ms when parallelized on 6 CPU threads.}
\label{fig:adam}
\end{figure}

As designed, the GO3 test-cases are generally intractable: the largest case, containing 23,643, buses, has 26,870 separate contingencies and 33,739 switchable lines/\newtxt{transformers} \textit{at each time step}. Reliably solving DC unit commitment (DCUC) on such a system is challenging in itself, much less ACUC. In order to overcome such levels of computational intractability, this paper, and the \btt{QuasiGrad} solver which it proposes, leverages techniques designed to solve large scale optimization problems from another community: machine learning (ML). The popularized GPT-3 model, for example, contains 175 billion tunable model parameters (i.e., optimizable \textit{decision variables})~\cite{brown2020language}. A plethora of gradient-based optimization tools have come from the ML community for solving such problems~\cite{choi2019empirical}, but Adam has emerged as the clear dominant solver. GPT-3 was trained with Adam~\cite{brown2020language}, as were other large commercial models.

While Adam has been a successful tool for solving large-scale nonlinear programming (NLP) problems, it has also recently been used to solve sub-problems in massive Mixed-Integer Linear Programming (MILP) Branch-and-Bound (BaB) problems. The $\alpha,\beta$-CROWN solver~\cite{Wang:2021}, which has won the most recent International Verification of Neural Networks Competitions (VNN-COMPs)~\cite{müller2023international}, uses a GPU-accelerated Adam solver to verify the performance of extremely large neural networks (whose verification problems often reformulate directly into MILPs). Notably, $\alpha,\beta$-CROWN was able to win these highly competitive competitions by carefully leveraging the computational might of modern GPUs along with the inherent \textit{parallelizability} of Adam. Like other gradient-based solvers, Adam requires a massive number \newtxt{of} derivative computations -- each of these ``backpropagations" can be efficiently computed in parallel, greatly accelerating the iterations.

Building on these successes, this paper introduces the \btt{QuasiGrad} solver for solving large scale power system optimization problems (specifically, the one formulated for GO3). The core numerical workhorse under \btt{QuasiGrad}'s hood is Adam. Despite its prowess, Adam needs help: \newtxt{to get this help}, \btt{QuasiGrad} leverages a number of other innovations to aid in solving the large-scale reserve and security constrained ACUC market clearing problem. Four of these contributions are summarized before:
\begin{enumerate}
\item We reformulate, and explicitly backpropagate through, the GO3 reserve + security constrained ACUC problem.
\item Using a preconditioned conjugate gradient (Kyrlov subspace) method, we stochastically select, numerically solve, and backpropagate through, parallelized security constraints. The computed gradients are passed to Adam.
\item We design a series of parallelized projection methodologies, including a guaranteed feasible ramp-constrained power flow solver, in order to exploit parallel computational resources and accelerate the convergence of Adam.
\item We release \btt{QuasiGrad.jl}, an open-source package available in the Julia ecosystem~\newtxt{~\cite{QuasiGrad_github}}.
\end{enumerate}

In Sec.~\ref{sec_problem}, we introduce the Mixed-Integer NLP (MINLP) proposed in GO3, and we transform the problem to make it amenable for gradient-based solvers. In Sec~\ref{sec_qg}, we propose the full \btt{QuasiGrad} solver, and in Sec.~\ref{sec_tests}, we provide simulated test results. Conclusions are presented in Sec.~\ref{conclusion}.

\section{Problem Formulation}\label{sec_problem}
This section introduces the MINLP designed by the GO3 planning team. This MINLP is then transformed into a formulation which is amenable for gradient based solvers (i.e., Adam) to directly interact with. Regarding notation, lower case variables are scalars (e.g., $x$), bold lower case variables are vectors (e.g., $\bm x$), and upper case variables are generally matrices (e.g., $X$). In all cases, we closely follow the notation prescribed in the official GO3 problem formulation~\cite{GO3}.

\subsection{Motivation for Reformulation}
Training problems in ML are often formulated as
\begin{align}\label{eq: ml_loss}
\min_{\bm{x}}\;\,\mathcal{L}(\bm{x}),
\end{align}
where $\mathcal{L}(\bm{x})$ is the canonical ``loss" function relating input/output data mappings, and $\bm{x}$ is an unconstrained vector of model parameters. The \btt{QuasiGrad} solver is designed around the idea of transforming the GO3 MINLP problem into a form which approximates \eqref{eq: ml_loss}. As formulated in~\cite{GO3}, however, The GO3 MINLP contains three challenging complications:
\begin{enumerate}
\item equality constraints,
\item inequality constraints,
\item integer variables.
\end{enumerate}
While directly penalizing these constraints/integrality requirements and pushing them into the loss function may seem to be an obvious solution, care must be taken to ensure the penalization procedure does not introduce unnecessary, and potentially intractable (in the case of contingency violations), loss error. For example, consider the following NLP:
\begin{subequations}\label{eq:lp_ex}
\begin{align}
\min_{\bm{x},\bm{y}}\quad & f(\bm{x},\bm{y})\\
{\rm s.t.}\quad & A\bm{x}=\bm{y}.
\end{align}
\end{subequations}
The \textit{best} reformulation substitutes the equality $\bm{y}\leftarrow A\bm{x}$ directly, such that $f(\bm{x},A\bm{x})$ is minimized independently. Naive constraint penalization on the other hand, via ${\mathcal L= }f(\bm{x},\bm{y})+\lambda\left\Vert A\bm{x}-\bm{y}\right\Vert$, is problematic for three reasons:
\begin{enumerate}
\item the ``$\bm y$'' variable is unnecessarily retained; 
\item the optimizer must expend implicit computational resource in approximating $A\bm{x}\approx\bm{y}$;
\item finally, and most subtly, assume a given numerical solution $\bm{x}^*$ to \eqref{eq:lp_ex} is evaluated by computing $f(\bm{x}^*,A\bm{x}^*)$. In this case, any effort spent by the optimizer to minimize the penalty term $\lambda\left\Vert A\bm{x}-\bm{y}\right\Vert$ was a ``numerical distraction", since it had no explicit effect on solution quality. In summary, a successful reformulation will ``find $\bm{x}$ and compute $\bm{y}$" rather than ``find $\bm{x}$ and find $\bm{y}$".
\end{enumerate}
In light of these observations, the following subsections rewrite the GO3 MINLP into a form which is similar to \eqref{eq: ml_loss} and, thus, amenable for gradient based solvers (i.e., Adam). In doing so, we eliminate all unnecessary intermediate variables (i.e., ``$\bm{y}$" terms, which we call ``auxiliary" variables), and we ensure the gradient solver expends its computational resource computing numerical variable values which actually influence solution quality (i.e., ``$\bm{x}$" terms, which we call ``basis" variables).

\subsection{MINLP Reformulation}\label{ss_reform}
We begin by stating the following MINLP, which represents an exact transformation\footnote{The only GO3 constraint which is \textit{not} explicitly captured via \eqref{eq: go3_minlp} is the synchronous network connectivity constraint. This constraint specifies that a given line switch cannot induce an electrical island within the AC network. The \btt{QuasiGrad} solver deals with this constraint heuristically.} of the GO3 security and reserve constrained ACUC problem~\cite{GO3}. This transformation writes all auxiliary variables $\bm{y}$ as an explicit function of basis variables $\bm{x}$. Notably, we have left the contingency constraint function, $\bm{h}_{\rm ctg}(\cdot)$, in place, since it will receive special consideration:
\begin{subequations}\label{eq: go3_minlp}
\begin{align}
\min_{\bm{x}_{d},\bm{x}_{c}}\quad & z^{{\rm ms}}(\bm{x}_{c},\bm{x}_{d},\bm{y}) + z^{\rm ctg}\\
{\rm s.t.}\quad & \bm{y}={\bm f}(\bm{x}_{c},\bm{x}_{d})\label{eq: y=fx}\\
 & \bm{0}=\bm{h}_{\rm ctg}(\bm{x}_{c},\bm{x}_{d},\bm{\theta}_{k})\\
 & A_{c}\bm{x}_{c}+A_{d}\bm{x}_{d}\ge{\bm 0}\\
 & \underline{\bm{x}}_{c}\le\bm{x}_{c}\le\overline{\bm{x}}_{c}\\
 & \underline{\bm{x}}_{d}\le\bm{x}_{d}\le\overline{\bm{x}}_{d}\\
 & \bm{x}_{d}\in\mathbb{Z}^{n_{d}}\label{eq: bb1}\\
 & \bm{x}_{c}\in\mathbb{R}^{n_{c}}\label{eq: bb2},
\end{align}
\end{subequations}
where $\bm{x}_{c}$ and $\bm{x}_{d}$ are vectors of continuous and discrete \textit{basis} variables (we denote ${\bm x}$ as the more general concatenation of ${\bm x}_c$ and ${\bm x}_d$), while $\bm{y}$ is a vector of auxiliary variables. We use the term ``basis" to denote the minimum set of variables needed to uniquely reconstruct a full GO3 solution (e.g., $\bm v$, $\bm \theta$, etc.); these are the same variables which are reported in the \textit{solution.json} file sent to the GO3 solution parser (referred to as ``Output data" in~\cite{GO3}). The basis variable sets $\Omega_c$ and $\Omega_d$, associated with $\bm{x}_{c}$ and $\bm{x}_{d}$,
are given as
\begin{align}
\Omega_c=\{ & v\cup\theta\cup\phi\cup\tau\cup p^{{\rm fr,dc}}\cup q^{{\rm fr,dc}}\cup q^{{\rm to,dc}}\cup p^{{\rm on}}\cup \nonumber\\
 & q\cup p^{{\rm rgu}}\cup p^{{\rm rgd}}\cup p^{{\rm scr}}\cup p^{{\rm nsc}}\cup p^{{\rm rru,on}}\cup\nonumber\\
 & p^{{\rm rru,off}}\cup p^{{\rm rrd,on}}\cup p^{{\rm rrd,off}}\cup q^{{\rm rqu}}\cup q^{{\rm rqd}}\}\label{eq: omega_c}\\
 \Omega_d=\{&u^{{\rm sh}}\cup u^{{\rm on}}\}.\label{eq: omega_d}
\end{align}
See~\cite{GO3} for definitions. The auxiliaries in $\bm{y}$, on the other hand, represent the large set of variables which can be directly eliminated from the problem formulation. In the reformulation \eqref{eq: go3_minlp}, we have carefully eliminated all such auxiliary variables. Our elimination uses the ${\rm max}(\cdot, 0)$, or ReLU, operator extensively; the ReLU is one of the foundational nonlinear activation functions used in modern ML, and gradient-based solvers have demonstrated a remarkable ability to optimize over functions which use it (to the apparent surprise of mathematicians~\cite{strang2019linear}).

Next, we present a series of four representative auxiliary variable reformulation examples; i.e., where we reformulate optimization constraints into to explicit functions à la \eqref{eq: y=fx}.

$\bullet$ \textit{Example 1 (slack variable elimination):}
The apparent power flow on a given line $s_{jt}$ is penalized if its magnitude is larger than flow limit $s_{j}^{\mathrm{max}}$. In~\cite{GO3}, this penalty $z_{jt}^{\mathrm{s}}$ is formulated using slack variable $s_{jt}^{+}$ (``fr" and ``to" sides are neglected for notational clarity):
\begin{subequations}
\begin{align}
0\leq & s_{jt}^{+}\\
z_{jt}^{\mathrm{s}}= & d_{t}c^{\mathrm{s}}s_{jt}^{+}\\
\sqrt{p_{jt}^{2}+q_{jt}^{2}}\leq & s_{j}^{\mathrm{max}}+s_{jt}^{+}.
\end{align}
\end{subequations}
Since line flows may be written as direct nonlinear functions of the basis variables, and since the slack value may be captured using a $\max$ operator, the penalty may be computed as an explicit function of basis variables $\bm x$:
\begin{align}
z_{jt}^{\mathrm{s}}=d_{t}c^{\mathrm{s}}\max((p_{jt}^{2}(\bm{x})+q_{jt}^{2}(\bm{x}))^{1/2}-s_{j}^{\mathrm{max}},0).
\end{align}
All slack variables are transformed in this way.\hfill\qedsymbol{}

$\bullet$ \textit{Example 2 (startup and shutdown variable elimination):}
Startup, shutdown, and on-off variables are linked via evolution equations~\cite{GO3}:
\begin{subequations}\label{eq: u_ev}
\begin{align}
u_{jt}^{\mathrm{su}}+u_{jt}^{\mathrm{sd}} & \le1\\
u_{jt}^{\mathrm{on}}-u_{j,t-1}^{\mathrm{on}} & =u_{jt}^{\mathrm{su}}-u_{jt}^{\mathrm{sd}}.
\end{align}
\end{subequations}
The startup and shutdown variables, however, are uniquely defined for a given on-off variable $u_{jt}^{\mathrm{on}}$ sequence. Therefore, \eqref{eq: u_ev} may be captured via the following explicit definitions for auxiliary startup and shutdown variables:
\begin{align}
u_{jt}^{\mathrm{su}} & \triangleq+\max\left(u_{jt}^{\mathrm{on}}-u_{j,t-1}^{\mathrm{on}},0\right)\\
u_{jt}^{\mathrm{sd}} & \triangleq-\min\left(u_{jt}^{\mathrm{on}}-u_{j,t-1}^{\mathrm{on}},0\right).
\end{align}
These auxiliary variables are then plugged in for a variety of uses (e.g., startup state calculations, shutdown costs, etc). \hfill\qedsymbol{}

$\bullet$ \textit{Example 3 (device cost curves):}
Time-dependent device costs $z_{jt}^{\mathrm{en}}$ are modeled in~\cite{GO3} via piecewise linear convex (or concave) cost (or value) functions:
\begin{subequations}\label{eq: gen_cost}
\begin{align} 
 & 0\leq p_{jtm}\leq p_{jtm}^{\max},\;\forall t\in T,j\in J^{\mathrm{pr},\mathrm{cs}},m\in M_{jt}\\
 & p_{jt}=\sum_{m\in M_{jt}}p_{jtm},\;\forall t\in T,j\in J^{\mathrm{pr},\mathrm{cs}}\\
 & z_{jt}^{\mathrm{en}}=d_{t}\sum_{m\in M_{jt}}c_{jtm}^{\mathrm{en}}p_{jtm},\;\forall t\in T,j\in J^{\mathrm{pr},\mathrm{cs}}.
\end{align}
\end{subequations}
By instead defining a cumulative block size $p_{jtm_{L}}^{{\rm cum},\max}$ as
\begin{align}
p_{jtm_{L}}^{{\rm cum},\max}=\sum_{l=1}^{L}p_{jtm_{l}}^{\max},
\end{align}
energy cost may be explicitly computed as
{\small
\begin{align}\label{eq: costs}
z_{jt}^{\mathrm{en}}=d_{t}\!\!\sum_{l=1}^{|M_{jt}|}c_{jtm_{l}}^{\mathrm{en}}\max\!\left(\min\left(p_{jt}-p_{jtm_{L=l-1}}^{{\rm cum},\max},p_{jtm_{l}}\!\right)\!,0\right)\!,
\end{align}}
$\!\!$where $p_{jtm_{L=0}}^{{\rm cum},\max}=0$. The $\max(\min(\cdot))$ formulation sums the length of a bid block, times its marginal cost, until a bid block component exceeds the power production value $p_{jt}$ (at which point, 0 is added thereafter). \hfill\qedsymbol{}

$\bullet$ \textit{Example 4 (power balance):}
Active and reactive power imbalance expressions are computed in~\cite{GO3} as
{\small
\begin{align} 
 & p_{it}=\sum_{j\in J_{i}^{\mathrm{cs}}}p_{jt}\!+\!\sum_{j\in J_{i}^{\mathrm{sh}}}p_{jt}\!+\!\sum_{j\in J_{i}^{\mathrm{fr}}}p_{jt}^{\mathrm{fr}}\!+\!\sum_{j\in J_{i}^{\mathrm{to}}}p_{jt}^{\mathrm{to}}\!-\!\sum_{j\in J_{i}^{\mathrm{pr}}}p_{jt}\\
 & q_{it}=\sum_{j\in J_{i}^{\mathrm{cs}}}q_{jt}\!+\!\sum_{j\in J_{i}^{\mathrm{sh}}}q_{jt}\!+\!\sum_{j\in J_{i}^{\mathrm{fr}}}q_{jt}^{\mathrm{fr}}\!+\!\sum_{j\in J_{i}^{\mathrm{to}}}q_{jt}^{\mathrm{to}}\!-\!\sum_{j\in J_{i}^{\mathrm{pr}}}q_{jt}.
\end{align}}
Mismatch penalties, $\forall t\in T,i\in I$, are then computed via $z_{it}^{\mathrm{p}} =d_{t}c^{\mathrm{p}}p_{it}^{+}$ and $z_{it}^{\mathrm{q}} =d_{t}c^{\mathrm{q}}q_{it}^{+}$, where slack inequalities $p_{it}^+\ge p_{it}$, $p_{it}^+\ge -p_{it}$, $q_{it}^+\ge q_{it}$, and $q_{it}^+\ge q_{it}$ are additionally enforced. To transform this expression, we take absolute value:
\begin{align}
z_{it}^{\mathrm{p}} & =d_{t}c^{\mathrm{p}}|p_{it}|\label{eq: p_bal}\\
z_{it}^{\mathrm{q}} & =d_{t}c^{\mathrm{q}}|q_{it}|,\label{eq: q_bal}
\end{align}
where $p_{it}$ and $q_{it}$ come directly from the mismatch equations defined above. Since $c^{\mathrm{p}}$ and $c^{\mathrm{q}}$ are very large penalization constants, a tightening ``soft-abs" function will be used in the final \btt{QuasiGrad} formulation, rather than an abs function.\hfill\qedsymbol{}

Through these transformations, along with a series of other similar ones (see SI material for this paper:~\cite{SI}), we are able to write a market surplus function which is an explicit function of the 21 basis variables identified in \eqref{eq: omega_c}-\eqref{eq: omega_d}. \newtxt{This process was implemented manually, and to the author's knowledge, no generalized code base or software libraries exist which can automatically perform said transformation. Recent tools, e.g., in~\cite{Shrirang:2023}, have used ReLU operators to penalize constraint violations, but the direct application of ReLU-based penalizations will not, e.g., eliminate unnecessary slack variables, or efficiently capture block-bid transformations as in \eqref{eq: costs}.}

\subsection{Further refinements: constraint penalization, variable clipping, and integer relaxation}
With the isolation of auxiliary variables, the MINLP of \eqref{eq: go3_minlp} can have its ($i$) auxiliary variables eliminated, ($ii$) linear constraints penalized with a soft-ReLU\footnote{The soft-abs function applied to scalar $x$ is defined via $|x|_{\texttt s} \triangleq \sqrt{x^{2}+\epsilon^{2}}$. The soft-ReLU function applied to scalar $x$ is $\sigma_{\texttt s}(x) \triangleq \sqrt{\max(x,0)^{2}+\epsilon^{2}}$.}, ($iii$) integers relaxed, and ($iv$) basis variables ``clipped" into a rectangular bounding box $\mathcal B$ defined from constraints \eqref{eq: bb1}-\eqref{eq: bb2}. The updated NLP formulation is given as:
\begin{subequations}\label{eq: nlp_updated}
\begin{align}
\min_{\bm{x}_{d},\bm{x}_{c}\in\mathcal{B}}\quad & z^{{\rm ms}}(\bm{x}_{c},\bm{x}_{d},\bm{f}(\bm{x}_{c},\bm{x}_{d}))+z^{{\rm ctg}}\nonumber\\
 & \qquad\quad\quad+\rho\cdot\sigma_{s}\left(A_{c}\bm{x}_{c}+A_{d}\bm{x}_{d}\right)\label{eq: nlp_no_g}\\
{\rm s.t.}\quad & \bm{0}=\bm{h}_{{\rm ctg}}(\bm{x}_{c},\bm{x}_{d},\bm{\theta}_{k}).\label{eq: g_ctg}
\end{align}
\end{subequations}
\textit{Integer variables:} The general strategy of the \btt{QuasiGrad} solver for dealing with integers goes as follows: ($i$) solve NLP \eqref{eq: nlp_updated} to some degree of accuracy, ($ii$) project the relaxed integers into the feasible space (see next subsection), ($iii$) permanently fix a subset of the integers whose projected values were closest to their relaxed values, adding them to set $\mathcal F$, and ($iv$) repeat until all integers are fixed to feasible values. This sort of \textit{batch rounding} procedure is a highly useful heuristic, but it can easily be replaced with a more systematic integer search (as with $\alpha,\beta$-CROWN, which performs a complete BaB search routine with Adam as the subproblem solver). This procedure is generally inspired by the Iterative Batch Rounding (IBR) routine used by the GravityX team in GO1 and GO2~\cite{GO2_review}. Our general strategy is outlined in Alg.~\ref{algo:int}, where $n_b$ is the total number of binaries. As the Adam solver iterates and the wall clock time increases, the soft-ReLU function in \eqref{eq: nlp_no_g} increasingly penalizes constraint violations more strongly, similar to the soft-abs tightening in Fig.~\ref{fig:homotopy}.

\begin{algorithm}
\caption{\btt{QuasiGrad} Process for Fixing Integers}\label{algo:int}

{\small

\begin{algorithmic}[1]

\State ${\mathcal F} \leftarrow \emptyset$ (no fixed binaries)

\While{$|{\mathcal F}| < n_b$ (some binaries are not fixed)}

\State Solve continuous NLP \eqref{eq: nlp_updated} with binaries in ${\mathcal F}$ fixed

\State Project device binaries via Proj.~\ref{bin_proj} using parallelized solves

\State Add a fraction of the projected binaries to ${\mathcal F}$

\EndWhile {\bf end}

\end{algorithmic}}
\end{algorithm}

\subsection{Integer projection}
\begin{figure}
\centering
\includegraphics[width=0.75\columnwidth]{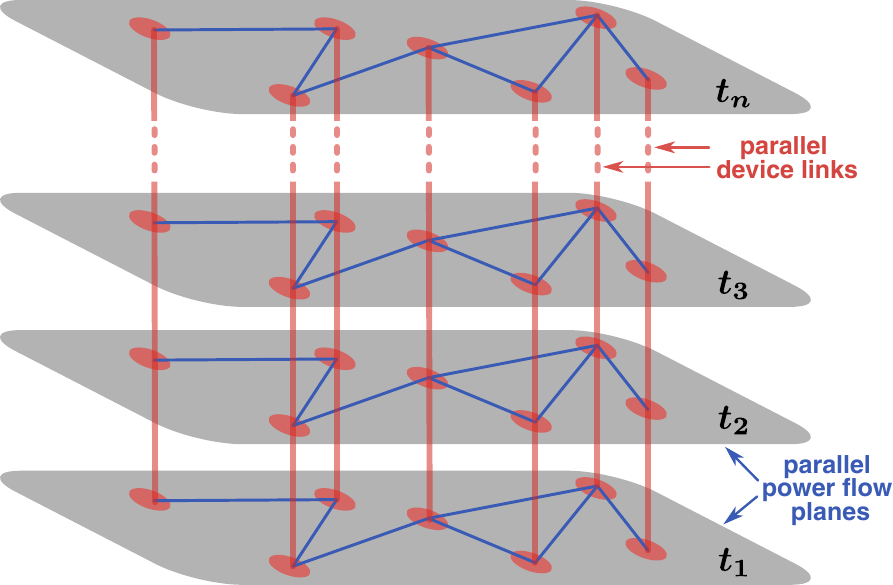}
\caption{The parallel nature of devices constraints and power flow constraints are portrayed. Devices can be projected feasible in parallel (via Proj.~\ref{bin_proj}), and power flow solves can be performed in parallel (via Proj.~\ref{proj_parallel_pf}).}\label{fig:parallel_devs_pfs}
\end{figure}

Adam does not enforce integrality constraints itself. Rather, \newtxt{in our approach, we rely} on the successive, highly-parallelized projection of all integer variables. The associated projection is given in Proj.~\ref{bin_proj}. This MILP projection enforces all ramp, reserve, headroom, and limit constraints for a given device -- producer (generator) or consumer (load). Notably, these projections  are executed in parallel via multi-threading (i.e., on each CPU thread, Gurobi is explicitly given a MILP to solve, each associated with a single device). The objective function tries to keep the decision variables $\bm{x}_{c}$, $\bm{x}_{d}$ as close to the continuous NLP solution $\bm{x}_{c}^{0}$, $\bm{x}_{d}^{0}$ as possible via penalizing deviations. The matrices $D_{c}^{{\rm g}_{i}}$, $D_{d}^{{\rm g}_{i}}$ are diagonal matrices which simply select the decision variables associated with device $i$. Finally, integer variables in $\mathcal{F}$ are fixed to their previously projected values and eliminated from a given projection. The parallel nature of device projections is illustrated by the parallel vertical red lines in Fig.~\ref{fig:parallel_devs_pfs}. Notably, a single device projection over, e.g., 18 time periods typically solves to $\sim\!0$ optimality gap \newtxt{via Branch-and-Bound search routine} in less than $50$ ms (often, even faster).

\begin{projection}[h]
\small
\caption{\hspace{-0.1cm}\textbf{:} Optimal Device Binary Projection [MILP]\\
\null$\quad\star$\textit{ \small parallelizable across each device}}\label{bin_proj}
\vspace{-0.25cm}
\begin{align}
\min_{\bm{x}_{c}\in {\mathbb R},\bm{x}_{d}\in {\mathbb Z}}\quad & \left\Vert D_{c}^{{\rm g}_{i}}\left(\bm{x}_{c}-\bm{x}_{c}^{0}\right)\right\Vert_1 +\left\Vert D_{d}^{{\rm g}_{i}}\left(\bm{x}_{d}-\bm{x}_{d}^{0}\right)\right\Vert_1 \nonumber \\
{\rm s.t.}\quad & \bm{x}_{d,i}=\bm{x}_{d,i}^{0},\;i\in\mathcal{F}\tag{\textbf{\textit{fixed binaries}}}\\
& \text{\cite[eqs. (48)-(58)]{GO3}}\tag{binary constraints}\\
& \text{\cite[eqs. (68)-(74)]{GO3}}\tag{ramp limits}\\
& \text{\cite[eq. (98)-(108)]{GO3}}\tag{reserve constraints}\\
& \text{\cite[eq. (109)-(118)]{GO3}}\tag{producer limits}\\
& \text{\cite[eq. (119)-(128)]{GO3}}\tag{consumer limits}
\end{align}
\vspace{-0.5cm}
\end{projection}

\subsection{Contingency gradients}
In subsection \ref{ss_reform}, we solved for and eliminated all auxiliary variables, but we left the contingency expression \eqref{eq: g_ctg}. Contingencies in GO3 are modeled via DC power flow solutions (in conjunction with nonlinear apparent power flow calculations). In order to eliminate a DC power flow expression $\bm{p}_{t}^{{\rm inj}}=Y_{k}\bm{\theta}_{tk}$, we would need to ($i$) solve it directly via $\bm{\theta}_{tk}=Y_{k}^{-1}\bm{p}_{t}^{{\rm inj}}$, ($ii$) use the phase angle solution to compute flow violations, and then ($iii$) push those violations up into the objective function \eqref{eq: nlp_no_g}. This approach is non-scalable, however, for two reasons:
\begin{itemize}
\item $Y_{k}^{-1}$ is generally a dense matrix, and it would consume very large amounts of memory to compute and store such matrices for each contingency/time;
\item every Adam iteration would require $n_t\times n_{\rm ctg}$ dense matrix-vector products, which would be untenable.
\end{itemize}
Instead, the approach we take can be summarized in three steps: at each Adam iteration, we ($i$) solve a subset of contingencies, ($ii$) backpropagate through the ones with the worst violations, and then ($iii$) pass these computed gradients to Adam. Thus, Adam does not \textit{solve} contingencies, but it does feel the pressure from their efficiently computed \textit{gradients}. In the following, we propose efficient methods for solving a subset of contingencies at each Adam iteration and then backpropagating through them.

\subsubsection{Contingency evaluation}
As in~\cite{GO3}, GO3 contingency $k$ line flows are computed via
\begin{subequations}
\begin{align}
p_{jtk} & =-b_{j}^{\mathrm{sr}}u_{jt}^{\mathrm{on}}\left(\theta_{itk}-\theta_{i^{\prime}tk}-\phi_{jt}\right)\\
 & =\underbrace{-b_{j}^{\mathrm{sr}}u_{jt}^{\mathrm{on}}\left(\theta_{itk}-\theta_{i^{\prime}tk}\right)}_{f_{jtk}}+\underbrace{b_{j}^{\mathrm{sr}}u_{jt}^{\mathrm{on}}\phi_{jt}}_{b_{jt}}.
\end{align}
\end{subequations}
We vectorize $f_{jtk}$, $b_{jt}$ across all lines and transformers into $\bm{f}_{tk}$, $\bm{b}_{t}$. We then use a signed incidence matrix $E$ to compute nodal injections (which are know by device injections):
\begin{subequations}
\begin{align}
\bm{p}_{t}^{{\rm inj}} & =E^{T}(\bm{f}_{tk}+\bm{b}_{t})\\
 & =E^{T}Y_{x}E\bm{\theta}_{tk}+E^{T}\bm{b}_{t}\\
 & =Y_{b}\bm{\theta}_{tk}+E^{T}\bm{b}_{t},
\end{align}
\end{subequations}
where $Y_{x}$ is a diagonal matrix of inverse line reactances. Deleting the reference bus (hat notation), reduced nodal angles $\hat{\bm{\theta}}_{tk}$ and contingency branch flows $\bm{p}_{tk}$ may be computed:
\begin{align}
\hat{\bm{\theta}}_{tk} & =\hat{Y}_{b}^{-1}\left(\hat{\bm{p}}_{t}^{{\rm inj}}-\hat{E}^{T}\bm{b}_{t}\right)\label{eq: Yb_solve}\\
\bm{p}_{tk} & =Y_{x}\hat{E}\hat{\bm{\theta}}_{tk}.\label{eq: p_flow}
\end{align}
Once computed, $\bm{p}_{tk}$ is used in conjunction with reactive power line flows to compute apparent power branch overloads. The main computational task in this process is solving the linear system in \eqref{eq: Yb_solve}. Here, we exploit a preconditioned conjugate gradient ($\texttt{pcg}$) solver~\cite{cg}. This solver is a Krylov subspace method which \textit{iteratively} (rather than recursively, in the case of Gaussian elimination) approximates a linear system solution. Since the base-case DC admittance matrix $\hat{Y}_{b}$ is static for a given set of line switches, we use a constant Limited memory LDL (\texttt{LLDL}) factorization as a preconditioner $\hat P$. This greatly accelerates $\texttt{pcg}$ convergence:
\begin{align}
\hat{P} & \leftarrow\texttt{LLDL}\left(\hat{E}^{T}Y_{x}\hat{E}\right)\\
\hat{\bm{\theta}}_{tb} & \approx\texttt{pcg}(\hat{\bm{p}}_{t}^{{\rm inj}}-\hat{E}^{T}\bm{b}_{t},\hat{Y}_{b},\hat{P}, \epsilon_{\rm pcg}),\label{eq: Y_base}
\end{align}
where the $\texttt{pcg}$ function approximates the solution of \eqref{eq: Yb_solve} with preconditioner $\hat{P}$; it terminates when the provided error metric $\epsilon_{\rm pcg}$ is satisfied. Notably, we only solve \eqref{eq: Y_base} for the base-case at time $t$ (i.e., no contingency branches removed from the network yet). Using a low-rank update procedure recently pioneered in~\cite{Holzer:SMW}, we then \textit{rank-1 correct} to solve for each contingency solution. This is motivated by the fact that the admittance matrix of a given contingency is only ``rank-1 away" from the base-case admittance matrix. To show this, let $Y_{k}$ be an almost-empty matrix with a single, nonzero entry; this entry is placed on the diagonal element associated with the single line that is removed in a contingency, and its value is the negative admittance of that line. Then, the relationship between the base-case admittance $\hat{Y}_{b}$ and a given contingency admittance matrix $\hat{Y}_{b,k}$ is
\begin{align}
\text{base-case admittance: }\quad\hat{Y}_{b} & =\hat{E}^{T}Y_{x}\hat{E}\nonumber\\
\text{contingency admittance: }\;\hat{Y}_{b,k} & =\hat{E}^{T}Y_{x}\hat{E}+\!\!\underbrace{\hat{E}^{T}Y_{k}\hat{E}}_{{\bm v}_{k} {\bm v}_{k}^T\text{ (rank-1)}}\!\!\!\!.\nonumber
\end{align}
Thus, using the Sherman-Morrison-Woodbury (SMW) formula~\cite{Horn:1990,Holzer:SMW}, we may rank-1 correct a base-case nodal phase angle solution. Setting ${\bm c}_t\triangleq \hat{\bm{p}}_{t}^{{\rm inj}}-\hat {E}^{T}\bm{b}_{t}$, we have
\begin{subequations}
\begin{align}
(\hat{Y}_{b}+{\bm v}_{k}{\bm v}_{k}^{T})\hat{\bm{\theta}}_{tk} & =\bm{c}_{t}\\
\hat{\bm{\theta}}_{tk} & =\left(\hat{Y}_{b}^{-1}-\frac{\hat{Y}_{b}^{-1}{\bm v}_{k}{\bm v}_{k}^{T}\hat{Y}_{b}^{-1}}{1+{\bm v}_{k}^{T}\hat{Y}_{b}^{-1}{\bm v}_{k}}\right)\bm{c}_{t}\\
 & =\hat{\bm{\theta}}_{tb}-\underbrace{\hat{Y}_{b}^{-1}{\bm v}_{k}}_{{\bm u}_{k}}\underbrace{\frac{{\bm v}_{k}^{T}\hat{Y}_{b}^{-1}}{1+{\bm v}_{k}^{T}\hat{Y}_{b}^{-1}{\bm v}_{k}}}_{{\bm w}_{k}}\bm{c}_{t}\\
 & =\hat{\bm{\theta}}_{tb}-{\bm u}_{k}({\bm w}_{k}^{T}\bm{c}_{t}),\label{eq: rank_correct}
\end{align}
\end{subequations}
since $\hat{\bm{\theta}}_{tb}\approx\hat{Y}_{b}^{-1}\bm{c}_{t}$ via \eqref{eq: Y_base}. Thus, the DC power flow solution to a given contingency can be computed as the rank-1 correction to a single \texttt{pcg} solve. This rank-1 correction is quickly computed with one vector-vector inner product, once vector-scalar product, and one vector-vector subtraction. After calculating $\hat{\bm{\theta}}_{tk}$, contingency active power flows are computed via \eqref{eq: p_flow}, and then directional line penalties are computed via
\begin{align}
{\bm s}_{tk}^{{\rm fr},+} & =\max\{(\bm{p}_{tk}^{2}+(\bm{q}_{tk}^{{\rm fr}})^{2})^{\tfrac{1}{2}}-{\bm s}^{\max,\mathrm{ctg}},0\}\label{eq: sfr_pen}\\
{\bm s}_{tk}^{{\rm to},+}  & =\max\{(\bm{p}_{tk}^{2}+(\bm{q}_{tk}^{{\rm to}})^{2})^{\tfrac{1}{2}}-{\bm s}^{\max,\mathrm{ctg}},0\}\\
z_{tk}^{{\rm ctg}} & ={\bm 1}^T \left(d_{t}c^{{\rm s}} \max\left({\bm s}_{tk}^{{\rm fr},+},{\bm s}_{tk}^{{\rm to},+},0\right)\right).\label{eq: zctg}
\end{align}

\subsubsection{Contingency backpropagation}
Impactful contingencies with nonzero penalties are backpropagated through, i.e., we take the gradient of aggregated penalty scalar $z_{tk}^{{\rm ctg}}$ with respect to all relevant basis variables. Taking the gradients with respect to reactive power flows, which are direct functions of nodal voltage variables, is fairly straightforward and explained in the SI~\cite{SI}. Active power injection gradients are less trivial and involve three steps: differentiate $z_{tk}^{{\rm ctg}}$ with respect to active power flows $\bm{p}_{tk}$, differentiate $\bm{p}_{tk}$ with respect to injections, and then differentiate injections with respect to basis variables. 

We begin this process with an example: assume some scalar $z=f({\bm y})$ has gradient $\nabla_{{\bm y}}z={\bm d}$. If ${\bm y}=A{\bm x}$, then
\begin{align}\label{eq: ex_Ax}
\left.\begin{array}{r}
\nabla_{\bm{y}}z=\bm{d}\\
\nabla_{\bm{x}}[y_1,y_2,...,y_n] =A^T
\end{array}\right\} \Rightarrow\ensuremath{\nabla_{\bm{x}}z=A^{T}\bm{d}}
\end{align}
is a well known result. We exploit this in the following way: we write the contingency penalty as a function of branch flows: $z_{tk}^{{\rm ctg}}=f(\bm{p}_{tk})$ whose gradient is $\nabla_{\bm{p}_{tk}}z_{tk}^{{\rm ctg}}=\bm{d}_{k}$ (this gradient can be written by inspection of \eqref{eq: sfr_pen}-\eqref{eq: zctg}). Next, we write the linear mapping between flows, injections, and phase shifters:
\begin{align}
\bm{p}_{tk} & =Y_{x,k}\hat{E}\hat{Y}_{k}^{-1}\hat{\bm{p}}_{t}^{{\rm inj}}-Y_{x,k}\hat{E}\hat{Y}_{k}^{-1}\hat{E}^{T}\bm{b}_{t}.
\end{align}
Fully analogous to \eqref{eq: ex_Ax}, the gradient mappings between contingency penalties, nodal injections, and phase shifters are
\begin{subequations}
\begin{align}
\nabla_{\hat{\bm{p}}_{t}^{{\rm inj}}}z_{tk}^{{\rm ctg}} & =\left(Y_{x,k}\hat{E}\hat{Y}_{k}^{-1}\right)^{T}\bm{d}_{k}\\
 & =\hat{Y}_{k}^{-1}\hat{E}^{T}Y_{x,k}\bm{d}_{k}\label{eq: 2nd_lin_sys}\\
\nabla_{\bm{b}_{t}}z_{tk}^{{\rm ctg}} & =-\left(Y_{x,k}\hat{E}\hat{Y}_{k}^{-1}\hat{E}^{T}\right)^{T}\bm{d}_{k}\\
 & =-\hat{E}\left(\nabla_{\hat{\bm{p}}_{t}^{{\rm inj}}}z_{tk}^{{\rm ctg}}\right).\label{eq: 2nd_lin_sys_phase}
\end{align}
\end{subequations}
Thus, in \eqref{eq: 2nd_lin_sys}, we are required to solve \textit{yet another} linear system. This is to be expected, since the contingency \textit{evaluation} \eqref{eq: Y_base} linear system solve incorporated no nonlinearity (it is simply a DC power flow solve), while the second linear system solve backpropagates through (i.e., takes sensitivity to) a number of nonlinear functions in \eqref{eq: sfr_pen}-\eqref{eq: zctg}). To solve \eqref{eq: 2nd_lin_sys}, we again use low-rank corrections to a \texttt{pcg} base-case solve.

There is one final, non-obvious step in the backpropagation. To take this step, we note that $\nabla_{\hat{\bm{p}}_{t}^{{\rm inj}}}\bm{p}_{tk}=(Y_{x,k}E\hat{Y}_{k}^{-1})^{T}$ is in fact an approximation -- not because of nonlinearity, but because of GO3 slack distribution rules. When taking a gradient, a power perturbation, say, on bus 1 must be \textit{uniformly} redistributed at all other buses according to \text{\cite[eq. (162)-(163)]{GO3}}. Thus, a power perturbation $\Delta p_{1}$ at bus 1 shows up like the following smaller perturbations at all other buses:
\begin{align}\label{eq: dp1}
\Delta p_{1}\rightarrow\left[\begin{array}{l}
p_{1}-\frac{\Delta p_{1}}{n}+\Delta p_{1}\\
p_{2}-\frac{\Delta p_{1}}{n}\\
\quad \vdots\\
p_{n}-\frac{\Delta p_{1}}{n}
\end{array}\right].
\end{align}
Thus, in the backpropagation, we need to correct for this effect. To do so, denote $A=Y_{x,k}\hat{E}\hat{Y}_{k}^{-1}$, where $\bm{p}_{tk}=A\hat{\bm{p}}_{t}^{{\rm inj}}$. By \eqref{eq: dp1}, each perturbation in power $\Delta p$ ``spreads out" across all powers, giving us the following scalar relation:
\begin{subequations}
\begin{align}
\Delta p_{tk,i} & =A_{i,j}\Delta\hat{p}_{t,j}^{{\rm inj}}-\frac{1}{n}\sum_{k}A_{i,k}\Delta\hat{p}_{t,j}^{{\rm inj}},\;\forall i,j\\
 & =\left(A_{i,j}-\frac{1}{n}A_{i}\bm{1}\right)\Delta\hat{p}_{t,j}^{{\rm inj}},\;\forall i,j.
\end{align}
\end{subequations}
Since these expressions hold $\forall i,j$, the matrix $A'$ which relates perturbations in injections and flows is given generally as
\begin{align}
A'=A-\frac{A{\bf 1}{\bf 1}^{T}}{n}=A\left(I-\frac{{\bf 1}{\bf 1}^{T}}{n}\right).
\end{align}
Directly updating the gradient in \eqref{eq: 2nd_lin_sys}, we have
\begin{subequations}
\begin{align}
\nabla_{\hat{\bm{p}}_{t}^{{\rm inj}}}z_{tk}^{{\rm ctg}} & =\left[Y_{x,k}\hat{E}\hat{Y}_{k}^{-1}\left(I-\frac{{\bf 1}{\bf 1}^{T}}{n}\right)\right]^{T}\bm{d}_{k}\\
 & =\left(I-\frac{{\bf 1}{\bf 1}^{T}}{n}\right)^{T}\underbrace{\hat{Y}_{k}^{-1}\hat{E}^{T}Y_{x,k}\bm{d}_{k}}_{\bm{\eta}}\\
 & =\bm{\eta}-\tfrac{1}{n}\sum\bm{\eta}.\label{eq: correction}
\end{align}
\end{subequations}
Thus, once we compute \eqref{eq: 2nd_lin_sys}, which yields $\bm{\eta}$, we correct its value via the surprisingly elegant \eqref{eq: correction}. Copious numerical test results confirmed the validity of this unexpected expression. \newtxt{Similar rank-1 flow corrections (e.g., more explicitly, to the PTDF matrix) have been observed in~\cite[eq. 9]{Huang:2023}).}

\subsubsection{Contingency backpropagation summary}
As Adam iterates, we maintain a running list of the most severe contingencies. At each Adam iteration, we evaluate contingencies in the top $X\%$ percentage of this list, along with a stochastic selection from the bottom percentage. All contingencies that have a $z_{tk}^{{\rm ctg}}$ score \eqref{eq: zctg} higher than a certain numerical threshold get backpropagated through; their gradients then get included in \eqref{eq: grads} and sent to Adam. Alg.~\ref{algo:ctg} summarizes this procedure.

\begin{algorithm}
\caption{Contingency solver}\label{algo:ctg}

{\small \textbf{Require:}
Set of worst contingencies ${\mathcal K}_t$, set of stochastically selected contingencies ${\mathcal S}_t$, \texttt{pcg} tolerance $\epsilon_{\rm pcg}$, backprop threshold $\zeta$

\begin{algorithmic}[1]

\For{$t \in T$} \Comment{\textbf{\textit{parallel}} loop over ACUC time periods}
\State \texttt{pcg} solve all base-case DC power flows via \eqref{eq: Y_base}
\EndFor {\bf end}

\For{$t \in T$}
\For{$k \in {\mathcal K}_t\cup {\mathcal S}_t$} \Comment{\textbf{\textit{parallel}} loop over contingencies}
\State Rank-1 correct base-case solutions via SMW: \eqref{eq: rank_correct}
\State Score ctg via \eqref{eq: zctg}
\If{$z_{tk}^{{\rm ctg}}>$ $\zeta$ threshold}
\State Solve backpropagation \eqref{eq: 2nd_lin_sys}-\eqref{eq: 2nd_lin_sys_phase} via \texttt{pcg} + SMW
\EndIf {\bf end}
\EndFor {\bf end}
\EndFor {\bf end}
\State Update contingency sets ${\mathcal K}_t$ and ${\mathcal S}_t$
\State \Return Approximated contingency gradients $\nabla_{\bm{x}}\bm{h}_{{\rm ctg}}$
\end{algorithmic}}
\end{algorithm}

\subsection{Implementation of the Adam solver}
We may now sum the contingency gradients $\nabla_{\bm{x}}\bm{h}_{{\rm ctg}}$ with the NLP objective function \eqref{eq: nlp_no_g} gradients:
\begin{tcolorbox}
\vspace{-11pt}
\begin{align}\label{eq: grads}
\bm{g}=\nabla_{\bm{x}}\left(z^{{\rm ms}}(\bm{x})+\rho\cdot\sigma_{s}\left(A\bm{x}\right)\right)+\nabla_{\bm{x}}\bm{h}_{{\rm ctg}},
\end{align}
\vspace{-18pt}
\end{tcolorbox}
$\!\!\!\!\!\!$where the shorthand $\bm x$ has been used to represent all basis variables. Eq.~\eqref{eq: grads} is the result of a \textit{backpropagation}. Notably, all gradient in~\eqref{eq: grads} are \textit{manually} computed in the \btt{QuasiGrad} solver source code, which required a fairly substantial effort. Backpropagation generates a cascade of derivatives which, via chain rule, connect the sensitivity of a loss function (or market surplus function) to a basis variable. For instance, backpropagation from basis variables which influence line flows ``$x_{\rm lf}$" to the market surplus function is given by
\begin{align}
\nabla_{x}z^{{\rm ms}}= & \nabla_{z^{{\rm base}}}z^{{\rm ms}}\cdot\nabla_{z_{t}^{\mathrm{t}}}z^{\mathrm{base}}\cdot\nabla_{z_{jt}^{\mathrm{s}}}z_{t}^{\mathrm{t}}\cdot\nabla_{s_{jt}^{+}}z_{jt}^{\mathrm{s}}\nonumber\\
&\cdot\nabla_{s_{jt}^{{\rm fr/to},+}}s_{jt}^{+}\cdot\nabla_{p/q_{jt}^{{\rm fr/to},+}}s_{jt}^{{\rm fr/to},+}\cdot\nabla_{x_{\rm lf}}p/q_{jt}^{{\rm fr/to},+},\nonumber\\
 &\qquad\qquad\qquad x_{\rm lf}\in\{v_{it},v_{i^{\prime}t},\theta_{it},\theta_{i^{\prime}t},\tau_{jt},\phi_{jt},u_{jt}^{\mathrm{on}}\}.\nonumber
\end{align}
Further details on gradient reformulation and backpropagation are provided in the SI~\cite{SI}. Importantly, computation of this gradient can exploit multi-threaded parallelism, which is the general key to Adam's success. Depending on the type of gradient needed, the \btt{QuasiGrad} solver parallelizes over ACUC time instance, network devices, or contingencies. One of the main tools for achieving such parallelism is Julia's \texttt{Threads.@threads} macro, which assigns the computational workload associated with a loop onto different available CPU threads. As a specific example, consider the gradient of a ``fr" line active power flow with respect to the ``to" side voltage (where $\bm{\delta}=\bm{\theta}_{t}^{{\rm fr}}-\bm{\theta}_{t}^{{\rm to}}-\bm{\phi}_{t}$):
\begin{align}
&\text{\textbf{for} } t \in T \nonumber \\
&\quad \quad \nabla_{\bm{v}_{t}^{{\rm to}}}\bm{p}_{t}^{\mathrm{fr}}=\left(-\bm{g}^{\mathrm{sr}}\cos\left(\bm{\delta}\right)-\bm{b}^{\mathrm{sr}}\sin\left(\bm{\delta}\right)\right)\bm{v}_{t}^{{\rm fr}}/\bm{\tau}_{t} \nonumber \\
&\text{\textbf{end} }\nonumber
\end{align}
Generally, a CPU will execute these gradient solves in series (one for each ACUC time instance). However, we may instruct the compiler to compute these gradients in parallel via
\begin{align}
&\text{\texttt{Threads.@threads} \textbf{for} } t \in T \nonumber \\
&\quad \quad \nabla_{\bm{v}_{t}^{{\rm to}}}\bm{p}_{t}^{\mathrm{fr}}=\left(-\bm{g}^{\mathrm{sr}}\cos\left(\bm{\delta}\right)-\bm{b}^{\mathrm{sr}}\sin\left(\bm{\delta}\right)\right)\bm{v}_{t}^{{\rm fr}}/\bm{\tau}_{t} \nonumber \\
&\text{\textbf{end} }\nonumber
\end{align}
Each doubling of CPU threads generally halves computational time. We note that there is no single ``right way" to multi-thread: an infinite variety of effective options exist.

After efficiently computing the gradients in \eqref{eq: grads}, we pass these gradients to a modified Adam solver~\cite{kingma2014adam}. Adam has been written about ad nauseam in the ML literature, so we provide little discussion of the Adam solver itself. \newtxt{At a high level, however, Adam uses first and second order moment estimates to implicitly track the curvature of a loss function landscape. Adam step sizes adaptively react to observed changes in the curvature, thus accelerating convergence towards some local minimum. Since Adam only needs gradient information to make decisions about step size and direction, variables can be updated in parallel. Furthermore, gradient calculations can computed concurrently, making the approach highly amenable to parallel computation.}

We feed the gradients from \eqref{eq: grads} into the modified Adam solver of Alg.~\ref{algo:adam}, which loops over basis variables and ACUC time instances (in parallel). After updating the Adam and basis variables states, basis variables are ``clipped" (i.e., projected) back into their feasible domain, as stated in Line 6. \newtxt{For example, a binary state $u$ is clipped into the range between 0 and 1 via $u\leftarrow\min(\max(u,0),1)$. A voltage state $v$ may be similarly clipped via $v\leftarrow\min(\max(v,{\underline v}),{\overline v})$.}

\begin{figure}
\centering
\includegraphics[width=0.75\columnwidth]{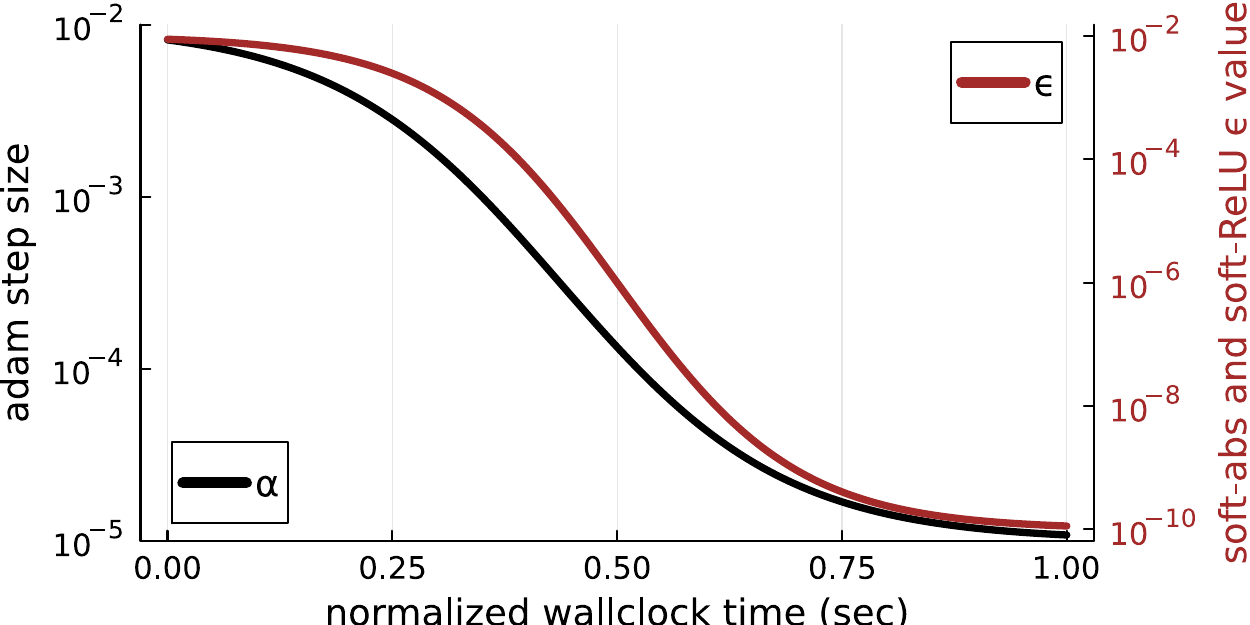}
\caption{Adam step size decay ($\alpha$, left) and soft-abs/soft-ReLU tightening ($\epsilon$, right). Step size decay ``leads" soft-abs/soft-ReLU tightening.}\label{fig:decay}
\end{figure}

\textit{Adam step size:} Practically speaking, one of the most important aspects of Alg.~\ref{algo:adam}
is setting the gradient descent step size $\alpha$. Since the GO3 clearing problems are time-limited (10, 120, and 240 minute limits for the real time, day-ahead, and week-ahead problems, respectively), we use wall-clock time to set the Adam step size: $\alpha_{\omega}( {\scriptstyle \StopWatchStart}=t_w\!)$, which is a function of basis variable $\omega$. Initially, Adam takes relatively large steps, but as time depletes, the steps sizes decay to very small values (in order to ``clean up" the solution). We use a reflected sigmoid function to set step size magnitude: 
\begin{align}
\text {normalize time: }\hat{t}_{w} & =2\frac{t_{w}-t_{0}}{t_{f}-t_{0}}-1\label{eq: norm_t}\\
\text {magnitude scale: }\beta & =\frac{e^{4\hat{t}_{w}}}{0.6+e^{4\hat{t}_{w}}}\label{eq: mag_scale}\\
\text {actual step size: }\alpha & =\alpha_{0}10^{\beta\cdot{\rm log10}(\frac{\alpha_{f}}{\alpha_{0}})}.\label{eq: step_size}
\end{align}
A representative step size decay curve, which exactly plots \eqref{eq: norm_t}-\eqref{eq: step_size}, is given by the black curve in Fig.~\ref{fig:decay}.

\begin{algorithm}
\caption{Modified Adam Solver (Original Adam:~\cite{kingma2014adam})}\label{algo:adam}
{\small \textbf{Require:}
Adam decay parameters $\beta_1,\beta_2$, adam iteration index $i$, basis variable gradients $\bm{g}_{t,\omega}$, step size function $\alpha_{\omega}( {\scriptstyle \StopWatchStart}\!)$
\begin{algorithmic}[1]
\For{$\omega \in \Omega$} \Comment{loop over basis variables$\qquad$$\qquad$$\qquad\!$}
\For{$t \in T$} \Comment{\textbf{\textit{parallel}} loop over ACUC time periods}
\State $\bm{m}_{t,\omega} \leftarrow\beta_{1}\cdot\bm{m}_{t,\omega}+\left(1-\beta_{1}\right)\cdot\bm{g}_{t,\omega}$
\State $\bm{v}_{t,\omega} \leftarrow\beta_{2}\cdot\bm{v}_{t,\omega}+\left(1-\beta_{2}\right)\cdot\bm{g}_{t,\omega}^{2}$
\State $\bm{x}_{t,\omega} \leftarrow\bm{x}_{t,\omega}-\alpha_{\omega}( {\scriptstyle \StopWatchStart}\!)\cdot\left({\tfrac{\bm{m}_{t,\omega}}{1-\beta_{1}^{i}}+\epsilon}\right)\!\Big/\!\left({\sqrt{\tfrac{\bm{v}_{t,\omega}}{1-\beta_{2}^{i}}}+\epsilon}\right)$
\State $\bm{x}_{t,\omega}\leftarrow\min(\max(\bm{x}_{t,\omega},\underline{\bm{x}}_{t,\omega}),\overline{\bm{x}}_{t,\omega})$ \Comment{clip all states!}
\EndFor {\bf end}
\EndFor {\bf end}
\end{algorithmic}}
\end{algorithm}

\begin{figure}
\centering
\includegraphics[width=0.75\columnwidth]{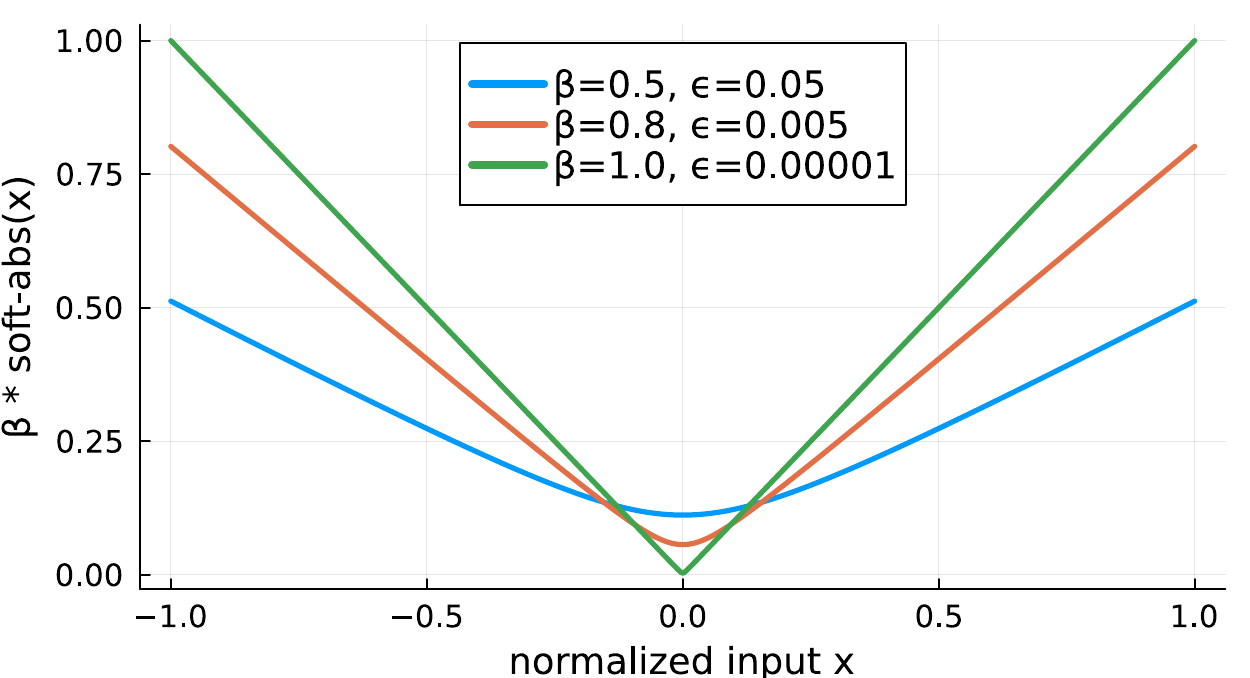}
\caption{Illustrated is the successive tightening of the soft-abs function $\beta\sqrt{x^{2}+\epsilon^{2}}$ as wall clock time increases. The value of epsilon is decreased in concordance with the red curve illustrated in Fig.~\ref{fig:decay}.}\label{fig:homotopy}
\end{figure}

\textit{Homotopic constraint penalization:} Certain gradients in \eqref{eq: grads} dominate other gradients by up to several orders of magnitude, essentially drowning out their contributions. In order to overcome this challenge, we first loosen these constraint penalties, and then we use a homotopy procedure to monotonically increase the penalization of these constraints as wall clock time increases. For example, we use a scaled, soft-abs function to penalize power balance error: $|x|_{s}=\beta\cdot\sqrt{x^{2}+\epsilon^{2}}$. The $\epsilon^{2}$ term decays in the same fashion as Adam step size \eqref{eq: norm_t}-\eqref{eq: step_size}, but over more orders of magnitude -- see the red curve in Fig.~\ref{fig:decay}, while $\beta$ increases linearly from $0.1$ to $1.0$. The effect of this homotopic penalization is demonstrated in Fig.~\ref{fig:homotopy}. Note that Adam is a first order method, so the upward shifting of these curves away from the origin is immaterial: only the \textit{gradient} of these curves is relevant. Homotopic constraint penalization is applied to power balance, branch flow, contingency, and ``penalized constraint" (see \eqref{eq: nlp_no_g}) penalties. Advanced homotopy methods have been highly successful in previous GO competitions~\cite{MCNAMARA2022108283}.

\section{\btt{QuasiGrad}}\label{sec_qg}
The previous section reformulated the GO3 MINLP in to form which Adam can interact with. While Adam is a powerful optimization tool, copious testing has found that Adam's effectiveness can be greatly enhanced if it is combined with other well-established, hyper-efficient numerical techniques (e.g., parallelized LP solvers). In this section, we review each of these tools, and we then we put them all together into a single, coherent solver (\btt{QuasiGrad}) with Adam at its center.

\subsection{Copper plate economic dispatch with LP relaxed binaries}
Since Adam is a local, gradient-based solver, it is highly influenced by its initialization. To initialize the \btt{QuasiGrad} solver, we first pose and solve a copper plate economic dispatch problem, where all integers are LP relaxed, and all contingency penalties and network variables are neglected. This LP\footnote{In some cases, the full LP is too large to be solved all at once (even on the GO3 evaluation platform, with 64 CPU cores and 256 GB of octa-channel DDR4-3200 memory). In these cases, we break the economic dispatch into parallelized sub-problems across shorter time periods; we then LP project device binaries to be feasible across all time (not shown here).}, which we classify as a projection, is given in Model \ref{model:cped}. Notably, the solution provides an excellent initialization for Adam, but the market surplus value also acts as a global upper bound on the MINLP -- this is useful for testing and bench-marking, as seen by the dashed orange line in Fig. \ref{fig:adam}.

\begin{projection}[h]
\small
\caption{\hspace{-0.1cm}\textbf{:} Copper Plate Economic Dispatch [LP]\\
\null$\quad\star$\textit{ \small \textbf{optionally} parallelizable across time instances}}
\label{model:cped}
\vspace{-0.25cm}
\begin{align}
\max_{\bm{x}_{c},\bm{x}_{d}}\quad & z^{\rm ms} \nonumber \\
{\rm s.t.}\quad & \text{\cite[eqs. (1)-(163)]{GO3}}\tag{nominal GO3 formulation}\\
& \textit{\textbf{neglect:}}\nonumber\\
& \quad\bullet \text{ shunts, contingencies, integers (LP relax)}\nonumber\\
& \quad\bullet \text{ all network variables (${\bm v}, {\bm \theta}, {\bm \tau}, {\bm \phi}$) and flow limits}\nonumber\\
& \textit{\textbf{impose:}}\nonumber\\
& \quad\bullet \sum_{j\in J^{{\rm pr}}}p_{jt}=\sum_{j\in J^{{\rm cs}}}p_{jt}+\sum_{j\in J^{{\rm dc}}}p_{jt}^{{\rm fr/to}},\; \forall t\tag{$p$ balance}\\
& \quad\bullet \sum_{j\in J^{{\rm pr}}}q_{jt}=\sum_{j\in J^{{\rm cs}}}q_{jt}+\sum_{j\in J^{{\rm dc}}}q_{jt}^{{\rm fr/to}},\; \forall t\tag{$q$ balance}
\end{align}
\vspace{-0.15cm}
\end{projection}

\subsection{Successively Linearized Power Flow Approximations}
Finding an AC power flow solution after solving the copper plate economic dispatch can be very challenging -- this is due to the fact that the economic dispatch often dispatches far more power than the network can physically accommodate (i.e., exceeding the maximum P-$\delta$ power transfer point of some lines). Adam can solve the resulting power flow, but schlepping large amounts of power across a network with gradient descent can be slower than ``hot starting" Adam with several linearized parallel power flow solves. We pose this linearized power flow problem in Proj.~\ref{proj_parallel_pf}, where 
\begin{itemize}
    \item $\bm{J}_{pv}^{\star}$, $\bm{J}_{p\theta}^{\star}$, $\bm{J}_{qv}^{\star}$, $\bm{J}_{q\theta}^{\star}$ are power balance sub-Jacobians;
    \item $\bm{J}_{sv}^{\star}$, $\bm{J}_{s\theta}^{\star}$ are apparent power flow sub-Jacobians;
    \item $\alpha\ge 1$ is an iteratively tightening flow constraint term;
    \item the objective function keeps all device injections as close to their initializations as possible via $\ell_2$ norm penalty. 
\end{itemize}

This projection also penalizes voltage perturbations (in order to regularize for convergence towards a consistent solution), and it regularizes for costs via $\gamma_5$ (i.e., pushing the solver towards cheaper power flow solutions). Notably, we use a quadratic objective function -- in testing, this was found to be much faster than a linear, $\ell_1$ norm penalizing function. The convex QPs of Proj.~\ref{proj_parallel_pf} are solved with Gurobi in parallel, as motivated by the parallel power flow planes in Fig. \ref{fig:parallel_devs_pfs}. 
\newtxt{These linearized projections are solved, and then iteratively re-solved, at the newly found solutions, which are used as new linearization points.} These successively linearized projections are \textit{not} run until convergence (i.e., $\Delta {\bm v} \approx \Delta {\bm \theta}\approx{\bm 0}$ is neither achieved nor desired), and a single QP can solve in less than 1 second for a network with several thousand buses.

\begin{projection}[h]
\small
\caption{\hspace{-0.1cm}\textbf{:} Regularized Power Balance Projection [QP]\\
\null$\quad\star$\textit{ \small parallelizable across each time instance}}
\label{proj_parallel_pf}
\vspace{-0.25cm}
\begin{align}\min_{\bm{x}_{c}}\quad & \gamma_{1}\left\Vert \bm{p}_{g}-\bm{p}_{g}^{0}\right\Vert_2 +\gamma_{2}\left\Vert \bm{q}_{g}-\bm{q}_{g}^{0}\right\Vert_2 +\gamma_{3}\Delta\bm{v}^{T}\Delta\bm{v}\nonumber\\
 & \qquad \quad +\gamma_{4}\Delta \bm{\theta}^{T}\Delta \bm{\theta}+\gamma_{5}\frac{c^{T}\bm{p}_{g}}{c^{T}\bm{p}_{g}^{0}}+\gamma_{6}\cdot\{\text{other regularizers}\}\nonumber\\
{\rm s.t.}\quad & p=p_{0}+\bm{J}_{pv}^{\star}\Delta v+\bm{J}_{p\theta}^{\star}\Delta\theta\tag{$p$ balance}\\
 & q=q_{0}+\bm{J}_{qv}^{\star}\Delta v+\bm{J}_{q\theta}^{\star}\Delta\theta\tag{$q$ balance}\\
 & s_{0}\bm{J}_{sv}^{\star}\Delta v+\bm{J}_{s\theta}^{\star}\Delta\theta\le\alpha\cdot\bm{s}^{\mathrm{max}}\tag{flow limits}\\
 & \underline{\bm{v}}\le\bm{v}+\Delta \bm{v}\le\overline{\bm{v}}\tag{voltage limits}\\
 & -72^{\circ}\le E\theta-\phi\le72^{\circ}\tag{angle limits}\\
 & p_{i}=\sum_{j\in J_{i}^{{\rm pr}}}p_{j}-\sum_{j\in J_{i}^{{\rm cs}}}p_{j}-\sum_{j\in J_{i}^{{\rm dc}}}p_{j}^{{\rm fr/to}}\tag{$p$ injection}\\
 & q_{i}=\sum_{j\in J_{i}^{{\rm pr}}}q_{j}-\sum_{j\in J_{i}^{{\rm cs}}}q_{j}-\sum_{j\in J_{i}^{{\rm dc}}}q_{j}^{{\rm fr/to}}\tag{$q$ injection}\\
& \text{\cite[eq. (109)-(118)]{GO3}}\tag{producer limits}\\
& \text{\cite[eq. (119)-(128)]{GO3}}\tag{consumer limits}\\
& \text{\cite[eq. (152)-(156)]{GO3}}\tag{dc line limits}
\end{align}
\vspace{-0.5cm}
\end{projection}

\subsection{Reserve variable cleanup}
The GO3 clearing problem is filled with reserve variables. In order to further help Adam, we often run a ``reserve variable cleanup" LP; this procedure very quickly tunes reserve variables via Proj.~\ref{model:reserve_cleanup} in order to minimize very costly reserve shortfalls in the cheapest way possible. These (parallelizable) LPs solve on the order of seconds on very large systems, but they save Adam a significant amount of computational effort.

\begin{projection}[h]
\small
\caption{\hspace{-0.1cm}\textbf{:} Reserve Product ``Cleanup" Projection [LP]\\
\null$\quad\star$\textit{ \small parallelizable across each time instance}}
\label{model:reserve_cleanup}
\vspace{-0.25cm}
\begin{align}\max_{\{\text{reserve variables}\}}\quad & -\sum_{j\in J^{\mathrm{pr},\mathrm{cs}}}\left(z_{jt}^{\mathrm{rgu}}+z_{jt}^{\mathrm{rgd}}+z_{jt}^{\mathrm{scr}}+z_{jt}^{\mathrm{nsc}}+z_{jt}^{\mathrm{rru}}\right)\nonumber\\
 & -\sum_{j\in J^{\mathrm{pr},\mathrm{cs}}}\left(z_{jt}^{\mathrm{rrd}}+z_{jt}^{\mathrm{qru}}+z_{jt}^{\mathrm{qrd}}\right)\nonumber\\
 & -\sum_{n\in N^{\mathrm{p}}}\left(z_{nt}^{\mathrm{rgu}}+z_{nt}^{\mathrm{rgd}}+z_{nt}^{\mathrm{scr}}+z_{nt}^{\mathrm{nsc}}+z_{nt}^{\mathrm{rru}}+z_{nt}^{\mathrm{rrd}}\right)\nonumber\\
 & -\sum_{n\in N^{\mathrm{q}}}\left(z_{nt}^{\mathrm{qru}}+z_{nt}^{\mathrm{qrd}}\right)\nonumber\\
{\rm s.t.}\quad & \text{\cite[eq. (20)-(47)]{GO3}}\tag{zonal reserve penalties}\\
& \text{\cite[eq. (80)-(108)]{GO3}}\tag{device reserve costs \& limits}\\
& \text{\cite[eq. (109)-(128)]{GO3}}\tag{device limits}
\end{align}
\vspace{-0.5cm}
\end{projection}

\subsection{Ramp-constrained power flow}
Adam is an excellent power flow solver, but its convergence towards an ``$\epsilon$ accurate" solution can be very slow, leading to unnecessary penalization in the final solution. In order to overcome this, the final step of the \btt{QuasiGrad} solver performs a ``ramp constrained" power flow projection. This projection is necessarily delicate, because it must respect the device limits -- if it does not, the device variables (i.e., $pq$ injections) have to be \textit{re-projected} feasible via Proj. \ref{bin_proj}, necessitating \textit{another} power flow solve; this cycle continues \textit{ad infinitum}, with no reason for convergence. Serial power flow solves which respect all future ramp constraints are extremely slow, both because they must be solved serially, and because they necessarily include linking constraints with all future power flow planes\footnote{For example, if the active power set-point of a device is to be altered at $t_1$, then we must ensure there exists a ramp-feasible power injection at $t_2$, but if the power at $t_2$ is being changed, then we must ensure there exists a ramp-feasible power injection at $t_3$, etc., until $t_f$.}. 

In order to overcome this challenge, we first consider $n$ parallel power flow problems
\begin{subequations}\label{eq: pfs}
\begin{align}
t_{1}:\; & \bm{f}_{1}(\bm{p}_{1},\bm{q}_{1},\bm{v}_{1},\bm{\theta}_{1})\\
t_{2}:\; & \bm{f}_{2}(\bm{p}_{1},\bm{q}_{1},\bm{v}_{1},\bm{\theta}_{1})\\
&\,\;\vdots\nonumber\\
t_{n}:\; & \bm{f}_{n}(\bm{p}_{n},\bm{q}_{n},\bm{v}_{n},\bm{\theta}_{n})
\end{align}
\end{subequations}
linked via ramp constraints 
\begin{subequations}\label{eq: ramps}
\begin{align}
t_{1}:\; & \bm{p}_{0}+\bm{d}_{1}^{{\rm rd}}\le\bm{p}_{1}\le\bm{p}_{0}+\bm{d}_{1}^{{\rm ru}}\label{eq: t1_ramp}\\
t_{2}:\; & \bm{p}_{1}+\bm{d}_{2}^{{\rm rd}}\le\bm{p}_{2}\le\bm{p}_{1}+\bm{d}_{2}^{{\rm ru}}\label{eq: t2_ramp}\\
&\,\;\vdots\nonumber\\
t_{n}:\; & \bm{p}_{n-1}+\bm{d}_{n}^{{\rm rd}}\le\bm{p}_{n}\le\bm{p}_{n-1}+\bm{d}_{n}^{{\rm ru}},
\end{align}
\end{subequations}
where $\bm{p}_0$, $\bm{d}_{i}^{{\rm ru}}$, and $\bm{d}_{i}^{{\rm rd}}$ are constants, and \eqref{eq: ramps} represent an exact transformation of the ramp limits \cite[eqs. (68)-(74)]{GO3} (once all binaries are frozen). 

\begin{figure}
\centering
\includegraphics[width=0.85\columnwidth]{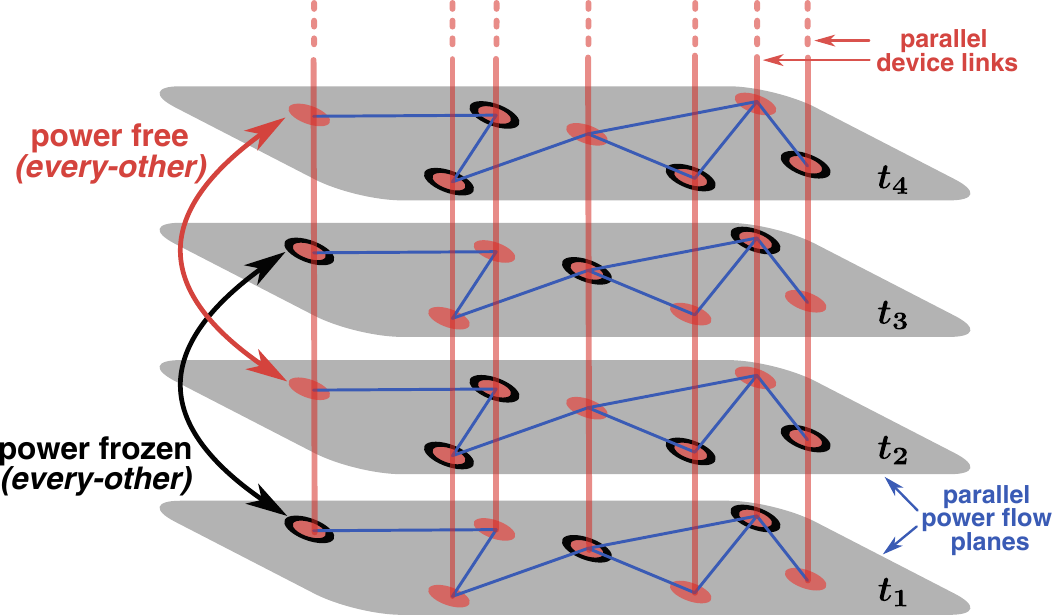}
\caption{Ramp-constrained power flow planes.}\label{fig:parallel_devs_pfs_ramp}
\end{figure}

In order to tractably solve \eqref{eq: pfs}-\eqref{eq: ramps}, we separate all network devices into two groups: the first group, termed group $a$, has its power injections frozen at $t_2$, $t_4$, $t_6$, etc., and second group, termed group $b$, has its power injections frozen at $t_1$, $t_3$, $t_5$, etc. Fig~\ref{fig:parallel_devs_pfs_ramp} demonstrates these alternatively frozen groupings. If a device $d$ power is frozen at time $t$, we say it belongs to set $\mathcal{F}_{d}^{t}$. Using these groupings, we pose the following associated constraints, where bracketed constraints are only enforced for the associated grouping ($a$ or $b$):
\begin{subequations}\label{eq: ramp_con_alt}
\begin{align}
t_{1}: & \left[\bm{p}_{0}+\bm{d}_{1}^{{\rm rd}}\le\bm{p}_{1}\le\bm{p}_{0}+\bm{d}_{1}^{{\rm ru}}\right]_{a},\;\left[\bm{p}_{1}=\bm{p}_{1}^{0}\right]_{b},\nonumber\\
 & \left[\bm{p}_{1}+\bm{d}_{2}^{{\rm rd}}\le\bm{p}_{2}\le\bm{p}_{1}+\bm{d}_{2}^{{\rm ru}}\right]_{a},\label{eq: t1_ramp_alt}\\
t_{2}: & \left[\bm{p}_{1}+\bm{d}_{2}^{{\rm rd}}\le\bm{p}_{2}\le\bm{p}_{1}+\bm{d}_{2}^{{\rm ru}}\right]_{b},\;\left[\bm{p}_{2}=\bm{p}_{2}^{0}\right]_{a},\nonumber\\
 & \left[\bm{p}_{2}+\bm{d}_{3}^{{\rm rd}}\le\bm{p}_{3}\le\bm{p}_{2}+\bm{d}_{3}^{{\rm ru}}\right]_{b},\label{eq: t2_ramp_alt}\\
t_{3}: & \left[\bm{p}_{2}+\bm{d}_{3}^{{\rm rd}}\le\bm{p}_{3}\le\bm{p}_{2}+\bm{d}_{3}^{{\rm ru}}\right]_{a},\;\left[\bm{p}_{3}=\bm{p}_{3}^{0}\right]_{b},\nonumber\\
 & \left[\bm{p}_{3}+\bm{d}_{4}^{{\rm rd}}\le\bm{p}_{4}\le\bm{p}_{3}+\bm{d}_{4}^{{\rm ru}}\right]_{a},\label{eq: t3_ramp_alt}\\
 & \;\;\,\, \vdots\nonumber
\end{align}
\end{subequations}
where devices are alternatively frozen and ramp constrained. Using this structure, the following result holds.
\begin{theorem}
Assume \eqref{eq: ramps} initially holds. By enforcing \eqref{eq: ramp_con_alt}, the power balance problems \eqref{eq: pfs} can be solved in parallel while maintaining ramp feasibility \eqref{eq: ramps} across all devices.
\begin{proof}
Going sequentially, \eqref{eq: t1_ramp_alt} directly implies \eqref{eq: t1_ramp}, since the initialized injections are ramp rate feasible. At $t_2$, \eqref{eq: t2_ramp_alt} implies \eqref{eq: t2_ramp}. This is because $[{\bm p}_2]_a$ was chosen in $t_1$ such that the $t_2$ ramp constraint for devices in group $a$ would be satisfied (these devices are frozen at $t_2$). The logic of choosing a device injection such that its ramp rate constraints are feasible at both the given and the following time step, and then freezing the device at the following time step, holds through to $t_n$.
\end{proof}
\end{theorem}
The parallelized ramp-constrained power flow projection is given in Proj.~\ref{model:ramp_constrained_pf}. \newtxt{As with Proj.~\ref{proj_parallel_pf}, this projection is iteratively re-solved as new linearization points are identified, thus driving the AC power balance constraints to satisfaction (the ``\textit{p} balance" and ``\textit{q} balance" constraints are linearized power mismatch expressions). Multi-period energy constraints are not enforced in this routine, since they inherently ``break" the parallelizability of the routine. In practice, however, this routine makes very small operational adjustments and rarely had any effect on the multi-period energy score.} This projection strikes a balance between speed (it is parallelizable) and solution quality, since at every time, one half of the devices can have their set-points updated in service of finding a minimally invasive power flow solution. This projection offers a useful solution to one of the most challenging problems faced by the author in solving GO3.

\begin{projection}[h]
\small
\caption{\hspace{-0.1cm}\textbf{:} Ramp-Constrained Power Flow at Time $t$ [QP]\\
\null$\quad\star$\textit{ \small parallelizable across each time instance}}
\label{model:ramp_constrained_pf}
\vspace{-0.25cm}
\begin{align}\min_{\bm{x}_{c}}\quad & \gamma_{1}\left\Vert \bm{p}_{g}-\bm{p}_{g}^{0}\right\Vert_2 +\gamma_{2}\left\Vert \bm{q}_{g}-\bm{q}_{g}^{0}\right\Vert_2 +\gamma_{3/4}\Delta\bm{v/\theta}^{T}\Delta\bm{v/\theta}\nonumber\\
{\rm s.t.}\quad & p=p_{0}+\bm{J}_{pv}^{\star}\Delta v+\bm{J}_{p\theta}^{\star}\Delta\theta\tag{$p$ balance}\\
 & q=q_{0}+\bm{J}_{qv}^{\star}\Delta v+\bm{J}_{q\theta}^{\star}\Delta\theta\tag{$q$ balance}\\
 & \underline{\bm{v}}\le\bm{v}+\Delta \bm{v}\le\overline{\bm{v}}\tag{voltage limits}\\
 & p_{i}=\sum_{j\in J_{i}^{{\rm pr}}}p_{j}-\sum_{j\in J_{i}^{{\rm cs}}}p_{j}-\sum_{j\in J_{i}^{{\rm dc}}}p_{j}^{{\rm fr/to}}\tag{$p$ injection}\\
 & q_{i}=\sum_{j\in J_{i}^{{\rm pr}}}q_{j}-\sum_{j\in J_{i}^{{\rm cs}}}q_{j}-\sum_{j\in J_{i}^{{\rm dc}}}q_{j}^{{\rm fr/to}}\tag{$q$ injection}\\
& \text{\cite[eqs. (68)-(74)]{GO3}}\tag{\textbf{ramp limits}}\\
& \text{\cite[eq. (109)-(118)]{GO3}}\tag{producer limits}\\
& \text{\cite[eq. (119)-(128)]{GO3}}\tag{consumer limits}\\
& \text{\cite[eq. (152)-(156)]{GO3}}\tag{dc line limits}\\
& p_{j}=p_{j}^{0},\;j\!\in\!\{\{J^{{\rm pr}}\!\cup\! J^{{\rm cs}}\}\cap\mathcal{F}_{d}^{t}\}\tag{\textbf{\textit{frozen subset via \eqref{eq: ramp_con_alt}}}}
\end{align}
\vspace{-0.5cm}
\end{projection}

\subsection{The \btt{QuasiGrad} solver}
We now algorithmically introduce the full \btt{QuasiGrad} solver. At a high level, the solver initializes a solution with an economic dispatch, and then it solves a series of NLPs while sequentially rounding and fixing binaries into feasible positions. The full algorithm is presented in Alg.~\ref{algo:qG}.

\begin{algorithm}
\caption{\btt{QuasiGrad}}\label{algo:qG}

{\small \textbf{Require:}
Total wall clock time ${\scriptstyle \StopWatchStart}$, total number of binaries $n_b$, number of binaries $n_b^+$ to freeze after every NLP (Adam) solve

\begin{algorithmic}[1]

\State ${\mathcal F} \leftarrow \emptyset$ (no fixed binaries)

\State Initialize with economic dispatch via Proj.~\ref{model:cped}

\While{$|{\mathcal F}| < n_b$ (some binaries are not fixed)}

\State Solve linearized power flow projection via Proj.~\ref{proj_parallel_pf}

\State Cleanup reserve variables via Proj.~\ref{model:reserve_cleanup}

\For{$t_w\in {\scriptstyle \StopWatchStart}_s$} \Comment{run adam for subset of wall clock time}

\State Evaluate and backpropagate objective \eqref{eq: g_ctg}

\State Solve and backpropagate contingencies via Alg.~\ref{algo:ctg}

\State Feed gradients \eqref{eq: grads} to Adam via Alg.~\ref{algo:adam}

\State Clip all states (Alg.~\ref{algo:adam}, line 6)

\EndFor{\bf end}

\State Project device binaries and variables feasible via Proj.~\ref{bin_proj}

\State Add $n_b^+$ binaries to frozen binary 
 set $\mathcal F$ based on projection

\EndWhile {\bf end}
\State Snap shunts

\State Solve linearized power flow projection via Proj.~\ref{proj_parallel_pf}

\State Cleanup reserve variables via Proj.~\ref{model:reserve_cleanup}

\State Run final Adam solve

\State Project device variables feasible via Proj.~\ref{bin_proj} (\textit{all binaries fixed})

\State Solve ramp-constrained power flow via Proj.~\ref{model:ramp_constrained_pf}

\State Cleanup reserve variables via Proj.~\ref{model:reserve_cleanup}

\State \Return feasible continuous ${\bm x}_c$ and discrete ${\bm x}_d$ solution vectors
\end{algorithmic}}
\end{algorithm}

\section{Test Results}\label{sec_tests}
In this section, we present simulated test results collected from division 1 (real time market) of the \texttt{C3E3.1} GO3 test-case library. We the provide brief comments about the official GO3 test results, which are partially available (in a very aggregated form) at the following footnote\footnote{\url{https://gocompetition.energy.gov/challenges/challenge-3/Leaderboards/Event-4}}.

The \texttt{C3E3.1} dataset\footnote{\url{https://gocompetition.energy.gov/challenges/600650/datasets}} contains six division 1 (i.e., real time market with 18 time periods) test cases: the 617-, 1576-, 4224-,
6049-, 6717-, 8316-, and 23643-bus systems. We provide local (i.e., laptop simulated) test results for a single \btt{QuasiGrad} solve of each of these cases, excluding the 23643-bus system, which cannot be solved locally due to memory constraints. For benchmarking, we compare the \btt{QuasiGrad} solution to the economic dispatch solution of Proj.~\ref{model:cped}, which is a global upper bound. All tests are run in Julia v1.10.0-beta1 on a Dell XPS with 16.0 GB of RAM. Julia is launched with access to 6 physical CPU threads for parallelization. Each solver terminates in under 10 minutes (600 seconds), as stipulated in GO3. All test results are confirmed feasible by the GO3 C3DataUtilities Python library\footnote{\url{https://github.com/GOCompetition/C3DataUtilities}}, and all reported scores are crosschecked against the C3DataUtilities solution scores.

Test results are reported in Table \ref{tab:tests}, which reports the \textit{gap} of the \btt{QuasiGrad} solution relative to the economic dispatch (i.e., $100.0 \times z^{\rm ms}/z^{\rm ed}$). The gaps are generally quite high, indicating that the \btt{QuasiGrad} solver was able to find high quality ACUC solutions within the 10 minute time allowance.

 \begin{table}
   \caption{C3E3.1 Division 1 Test Results} 
   \label{tab:tests}
   \small
   \centering
   \scalebox{0.9}{
   \begin{tabular}{c|cccccc}
   \toprule\toprule
   \textbf{testcase} & \textbf{617} & \textbf{1576} & \textbf{4224} & \textbf{6049}  & \textbf{6717} & \textbf{8316}\\
   \midrule 
   $z^{\rm ms}$ & 4.52e7   & 9.96e7  & 8.95e7   & 8.45e7  & 1.34e8 & 1.01e9  \\
   $z^{\rm ed}$ & 4.54e7   & 1.02e8  & 9.20e7   & 1.08e8  & 1.37e8 & 1.16e9  \\
   \textbf{gap}  & \textbf{99.8\%}   & \textbf{98.1\%}  &\textbf{97.3\%}   & \textbf{78.3\%}  & \textbf{97.7\%} & \textbf{87.0\%} \\
   \midrule 
   $z^{\rm base}$ & 4.53e7 & 1.00e8   & 8.96e7    & 8.45e7 & 1.34e8 & 1.01e9 \\
   $z^{\rm t}$    & 4.53e7 & 1.00e8   & 8.96e7    & 8.45e7 & 1.34e8 & 1.01e9 \\
   \midrule 
   & \multicolumn{6}{c}{\textbf{Relevant Penalty Breakdowns (\%):}}\\
   $z^{\rm en}$       & 98.6 & 62.9     & 90.8     & 51.7 & 66.0 & 89.8 \\
   $z^{\rm on/p/d}$ & 0.02 & 13.7     & 6.94     & 46.9 & 0.11 & 8.90 \\
   $z^{\rm ac}$       & 0.08 & 0.0      & 0.06     & 0.20 & 0.23 & 0.43 \\
   $z^{\rm xfm}$      & 0.0   & 0.0     & 0.49     & 0.17 & 0.0  & 0.04 \\
   $z^{\rm pq}$        & 0.0   & 0.0     & 0.0      & 0.0  & 0.0   & 0.0 \\
   $z^{\rm zonal}$    & 0.06  & 0.0     & 0.0      & 0.0  & 32.6  & 0.0 \\
   $z^{\rm ctg\text{-}min}$  & 1.18  & 23.1    & 1.21     & 0.56 & 1.08  & 0.55 \\
   $z^{\rm ctg\text{-}avg}$  & 0.05  & 0.32    & 0.52     & 0.38 & 0.01  & 0.28 \\
   \midrule 
\bottomrule
   \end{tabular}}
\end{table}

As of the submission of this manuscript, the full GO3 results have not yet been released. Overall, the \btt{QuasiGrad} solver performed well, but it was not a top performing algorithm. Across the 667 tests, \btt{QuasiGrad} found aggregated market surplus scores that were within $31\%$, $5\%$, and $44\%$, respectively, of the top performing team in the three market divisions, with 65 scores that were within the top 5 best. While there is significant room for improvement, the results demonstrate both the validity of this new approach, and the potential for it to be competitive with conventional approaches in the future.

\section{Conclusion}\label{conclusion}
This paper introduced an Adam-based solver, called \btt{QuasiGrad}  (summarized in Alg.~\ref{algo:qG}), capable of solving large-scale, reserve and security ACUC problems. The solver, which is released publicly as the Julia package \btt{QuasiGrad.jl}~\newtxt{~\cite{QuasiGrad_github}}, efficiently parallelizes backpropagation and variable projection processes, making efficient use of parallel computing hardware. The solver is able to find high quality solutions to large-scale problems in short periods of time, and by design, the approach is hyper-scalable. Due to its ability to efficiently parallelize, \btt{QuasiGrad} runs monotonically and predictably faster when it is given monotonically more computational resources. Future work will seek to test the \btt{QuasiGrad} solver on GPU hardware (which was not part of the GO3 competition). Planned follow-on work will provide deeper testing analysis and a more thorough investigation into the specific benefits of the innovations proposed in this paper. \newtxt{Furthermore, future directions should investigate the capacity for \btt{QuasiGrad} to help train physics-informed machine learning models, such as the Lagrange multiplier penalty-based learning models used in, e.g.,~\cite{Terrence:2023}.}

\section{Acknowledgements}\label{Ack}
The author gratefully acknowledges support and helpful feedback from Spyros Chatzivaseiliadis, Daniel Molzahn, Amrit Pandey, Mads Almassalkhi, and the GO3 organizers (especially Steven Elbert, Arun Veeramany, and Jesse Holzer). Finally, the author would like to thank Katherine and Rosemary Chevalier for their continued support throughout the development of QuasiGrad.

\appendices

{\section{}\label{AppA}}
\btt{QuasiGrad} was developed in \btt{Julia} \btt{v1.10.0}. All functions are type stable, and all memory is pre-allocated, thus minimizing the amount of garbage collection. Parallelization is achieved through \btt{Threads.@threads} and \btt{polyester.@batch}. \btt{LoopVectorization.jl}, and its macro \btt{@tturbo}, are used extensively to accelerate computations. The preconditioned conjugate gradient solver, \btt{cg!}, is called from \btt{IterativeSolvers.jl}. The \btt{lldl} function from  \btt{Preconditioners.jl} is used to build the limited memory LDL$^T$ preconditioner for contingency solving. \btt{JuMP.jl} and \btt{Gurobi.jl} are used to formulate and solve all optimizations (LPs, MILPs, and QPs). Gurobi 11 (and an associated academic license) was used.

\bibliographystyle{IEEEtran}
\bibliography{references}

\begin{thebibliography}{10}
\providecommand{\url}[1]{#1}
\csname url@samestyle\endcsname
\providecommand{\newblock}{\relax}
\providecommand{\bibinfo}[2]{#2}
\providecommand{\BIBentrySTDinterwordspacing}{\spaceskip=0pt\relax}
\providecommand{\BIBentryALTinterwordstretchfactor}{4}
\providecommand{\BIBentryALTinterwordspacing}{\spaceskip=\fontdimen2\font plus
\BIBentryALTinterwordstretchfactor\fontdimen3\font minus \fontdimen4\font\relax}
\providecommand{\BIBforeignlanguage}[2]{{%
\expandafter\ifx\csname l@#1\endcsname\relax
\typeout{** WARNING: IEEEtran.bst: No hyphenation pattern has been}%
\typeout{** loaded for the language `#1'. Using the pattern for}%
\typeout{** the default language instead.}%
\else
\language=\csname l@#1\endcsname
\fi
#2}}
\providecommand{\BIBdecl}{\relax}
\BIBdecl

\bibitem{aravena2023recent}
I.~Aravena, D.~K. Molzahn \emph{et~al.}, ``Recent developments in security-constrained ac optimal power flow: Overview of challenge 1 in the arpa-e grid optimization competition,'' \emph{Operations Research}, 2023.

\bibitem{GO2_review}
F.~Safdarian, J.~Snodgrass \emph{et~al.}, ``Grid optimization competition on synthetic and industrial power systems,'' in \emph{2022 North American Power Symposium (NAPS)}, 2022, pp. 1--6.

\bibitem{GO2}
J.~Holzer, C.~Coffrin \emph{et~al.}, ``Grid optimization competition challenge 2 problem formulation,'' \url{https://gocompetition.energy.gov/sites/default/files/Challenge2_Problem_Formulation_20210531.pdf}.

\bibitem{hijazi2018gravity}
H.~Hijazi, G.~Wang, and C.~Coffrin, ``Gravity: A mathematical modeling language for optimization and machine learning,'' 2018.

\bibitem{brown2020language}
T.~Brown, B.~Mann, N.~Ryder, M.~Subbiah, J.~D. Kaplan, P.~Dhariwal, A.~Neelakantan, P.~Shyam, G.~Sastry, A.~Askell \emph{et~al.}, ``Language models are few-shot learners,'' \emph{Advances in neural information processing systems}, vol.~33, pp. 1877--1901, 2020.

\bibitem{choi2019empirical}
D.~Choi, C.~J. Shallue, Z.~Nado, J.~Lee, C.~J. Maddison, and G.~E. Dahl, ``On empirical comparisons of optimizers for deep learning,'' \emph{arXiv preprint arXiv:1910.05446}, 2019.

\bibitem{Wang:2021}
S.~{Wang}, H.~{Zhang}, K.~{Xu}, X.~{Lin}, S.~{Jana}, C.-J. {Hsieh}, and J.~{Zico Kolter}, ``{Beta-CROWN: Efficient Bound Propagation with Per-neuron Split Constraints for Complete and Incomplete Neural Network Robustness Verification},'' \emph{arXiv e-prints}, p. arXiv:2103.06624, Mar. 2021.

\bibitem{müller2023international}
M.~N. Müller, C.~Brix, S.~Bak, C.~Liu, and T.~T. Johnson, ``The third international verification of neural networks competition (vnn-comp 2022): Summary and results,'' 2023.

\bibitem{QuasiGrad_github}
\BIBentryALTinterwordspacing
S.~Chevalier, ``Quasigrad.jl,'' Feb. 2024. [Online]. Available: \url{https://github.com/SamChevalier/QuasiGrad.jl}
\BIBentrySTDinterwordspacing

\bibitem{GO3}
J.~Holzer, C.~Coffrin \emph{et~al.}, ``Grid optimization competition challenge 3 problem formulation,'' \url{https://gocompetition.energy.gov/sites/default/files/Challenge3_Problem_Formulation_20230126.pdf}.

\bibitem{strang2019linear}
G.~Strang, \emph{Linear algebra and learning from data}.\hskip 1em plus 0.5em minus 0.4em\relax SIAM, 2019.

\bibitem{SI}
S.~Chevalier, ``Supplementary information for a parallelized, adam-based solver for reserve and security constrained ac unit commitment,'' \url{https://samchevalier.github.io/docs/SI.pdf}.

\bibitem{Shrirang:2023}
S.~Abhyankar, J.~Drgoňa, A.~Tuor, and A.~August, ``Neuro-physical dynamic load modeling using differentiable parametric optimization,'' in \emph{2023 IEEE Power \& Energy Society General Meeting (PESGM)}, 2023, pp. 1--5.

\bibitem{cg}
M.~R. Hestenes, E.~Stiefel \emph{et~al.}, ``Methods of conjugate gradients for solving linear systems,'' \emph{Journal of research of the National Bureau of Standards}, vol.~49, no.~6, pp. 409--436, 1952.

\bibitem{Holzer:SMW}
J.~T. Holzer, Y.~Chen, Z.~Wu, F.~Pan, and A.~Veeramany, ``Fast simultaneous feasibility test for security constrained unit commitment,'' \emph{IEEE Transactions on Power Systems}, pp. 1--10, 2023.

\bibitem{Horn:1990}
R.~Horn and C.~Johnson, \emph{Matrix Analysis}.\hskip 1em plus 0.5em minus 0.4em\relax Cambridge University Press, 1990.

\bibitem{Huang:2023}
Y.~Huang, T.~Ding, C.~Mu, X.~Zhang, Y.~He, and M.~Shahidehpour, ``Distributed slack-bus based dc optimal power flow with transmission loss: A second-order cone programming approach and sufficient conditions,'' \emph{IEEE Transactions on Automation Science and Engineering}, pp. 1--13, 2023.

\bibitem{kingma2014adam}
D.~P. Kingma and J.~Ba, ``Adam: A method for stochastic optimization,'' \emph{arXiv preprint arXiv:1412.6980}, 2014.

\bibitem{MCNAMARA2022108283}
\BIBentryALTinterwordspacing
T.~McNamara, A.~Pandey, A.~Agarwal, and L.~Pileggi, ``Two-stage homotopy method to incorporate discrete control variables into ac-opf,'' \emph{Electric Power Systems Research}, vol. 212, p. 108283, 2022. [Online]. Available: \url{https://www.sciencedirect.com/science/article/pii/S0378779622004722}
\BIBentrySTDinterwordspacing

\bibitem{Terrence:2023}
T.~W.~K. Mak, M.~Chatzos, M.~Tanneau, and P.~V. Hentenryck, ``Learning regionally decentralized ac optimal power flows with admm,'' \emph{IEEE Transactions on Smart Grid}, vol.~14, no.~6, pp. 4863--4876, 2023.

\end{thebibliography}

\end{document}